\documentclass[12pt]{article}

\usepackage{scicite}

\usepackage{times}

\usepackage{amscd}
\usepackage{amsfonts}
\usepackage{amsmath}
\usepackage{amssymb}
\usepackage{amsthm}
\usepackage{amsxtra}
\usepackage{bm}
\usepackage{bbm}
\usepackage{braket}
\usepackage{chemarrow}
\usepackage{caption}
\usepackage{comment}
\usepackage{dcolumn}
\usepackage{epsfig}
\usepackage{hyperref}
\usepackage{verbatim}
\usepackage{latexsym}
\usepackage{revsymb}
\usepackage{yfonts}
\usepackage{ifthen}
\usepackage{graphicx}
\usepackage{subfigure}
\usepackage{multirow}
\usepackage{enumerate}
\usepackage{epsfig}
\usepackage{xcolor}
\usepackage{ulem}
\usepackage{enumitem}
\usepackage{gensymb}
\usepackage{epstopdf}
\usepackage{color}
\usepackage{makecell}
\usepackage{stackengine}
\usepackage{upgreek}
\usepackage{ragged2e}
\usepackage{lineno}
\usepackage{float}
\usepackage[english]{babel}
\usepackage[utf8]{inputenc}
\usepackage{fancyhdr}

\renewcommand{\emph}[1]{\textit{#1}}

\def\ket#1{\mathinner{|{#1}\rangle}}

\hypersetup{colorlinks=true, linkcolor=blue, citecolor=red, urlcolor=blue}

\newenvironment{sciabstract}{%
\begin{quote} \bf}
{\end{quote}}

\title{Experimental quantum Byzantine agreement on a three-user quantum network with integrated photonics}

\author
    {Xu Jing$^{1\dagger}$, Cheng Qian$^{1\dagger}$, Chen-Xun Weng$^{2\dagger}$, Bing-Hong Li$^{2\dagger}$, \\
	Zhe Chen$^{1}$, Chen-Quan Wang$^{3}$, Jie Tang$^{3}$, Xiao-Wen Gu$^{3}$, \\Yue-Chan Kong$^{3}$, Tang-Sheng Chen$^{3}$, Hua-Lei Yin$^{4,2\ast}$, \\Dong Jiang$^{5\ast}$, Bin Niu$^{3,2\ast}$, Liang-Liang Lu$^{1,2,3,6\ast}$\\
\\
\normalsize{$^{1}$Key Laboratory of Optoelectronic Technology of Jiangsu Province,}\\
\normalsize{School of Physical Science and Technology,}\\
\normalsize{Nanjing Normal University, Nanjing 210023, China.}\\
\normalsize{$^{2}$National Laboratory of Solid-State Microstructures and School of Physics, }\\
\normalsize{Nanjing University, Nanjing 210093, China.}\\
\normalsize{$^{3}$National Key Laboratory of Solid-State Microwave Devices and Circuits,}\\
\normalsize{Nanjing Chip Valley Industrial Technology Institute,}\\
\normalsize{Nanjing Electronic Devices Institute, Nanjing, 210016, China.}\\
\normalsize{$^{4}$Department of Physics and Key Laboratory of Quantum State}\\
\normalsize{Construction and Manipulation (Ministry of Education),}\\
\normalsize{Renmin University of China, Beijing 100872, China.}\\
\normalsize{$^{5}$School of Internet, Anhui University, Hefei 230039, China.}\\
\normalsize{$^{6}$Hefei National laboratory, Hefei 230088, China.}\\
\normalsize{$^\dagger$These authors contributed equally to this work.}\\
\normalsize{$^\ast$Corresponding author. E-mail: hlyin@ruc.edu.cn (H.-L. Yin);}\\
\normalsize{jiangd@nju.edu.cn (D. Jiang); niubin$\_$1@126.com (B. Niu);}\\
\normalsize{lianglianglu@nju.edu.cn (L.-L. Lu)}
}

\date{}

\begin{document} 
\begin{sloppypar}


\baselineskip24pt


\maketitle 
\pagestyle{plain}


\begin{sciabstract}
Quantum communication networks are crucial for both secure communication and cryptographic networked tasks. Building quantum communication networks in a scalable and cost-effective way is essential for their widespread adoption. Here, we establish a complete polarization entanglement-based fully connected network, which features an ultrabright integrated Bragg reflection waveguide quantum source, managed by an untrusted service provider, and a streamlined polarization analysis module, which requires only one single-photon detector for each user. We perform a continuously working quantum entanglement distribution and create correlated bit strings between users. Within the framework of one-time universal hashing, we provide the experimental implementation of source-independent quantum digital signatures using imperfect keys circumventing the necessity for private amplification. We further beat the 1/3 fault-tolerance bound in the Byzantine agreement, achieving unconditional security without relying on sophisticated techniques. Our results offer an affordable and practical route for addressing consensus challenges within the emerging quantum network landscape.
\end{sciabstract}



\section*{INTRODUCTION}
Quantum communication \cite{Bennett1984Quantum,Artur1991Quantum,Lo1999Unconditional,Gisin2002Quantum,Scarani2009Security,Xu2020Secure} is one of the most mature quantum technologies. It enables the generation of secure keys between distant parties, even under the surveillance of an eavesdropper with unlimited computing power. A quantum link comprises fiber or free-space optical channels, along with a specific type of communication protocol. For example, quantum teleportation allows the transmission of fragile quantum information between remote parties using previously shared entanglement and classical communication \cite{Bouwmeester1997Experimental}. Quantum secure direct communication, another important branch of quantum communication, has offered opportunities for directly sending secret information over secure quantum channels \cite{Deng2003Twostep,Sheng2022Onestep,Zhou2022Onestep}. Quantum key distribution (QKD) is the fastest growing fields in quantum communication and has been implemented on a great variety of platforms, such as long-distance fiber \cite{Wang2019Twinfield,Liu2023Experimental} and free-space links \cite{Liao2018Satellite,Yin2020Entanglementbased}. Scaling the standard two-user protocols to many users is essential for its large-scale adoption \cite{Chen2021Integrated,Long2022Evolutionary,Pan2024Evolution}. Up to this point, several QKD networks have undergone trials with trusted nodes 
\cite{Peev2009SECOQC,Stucki2011Longterm,Wang2014Field} or passive/active switching \cite{Tang2016Measurement,Frhlich2013AQA,Lim2008Broadband,Aktas2016Entanglement} at the cost of sacrificing security or functionality. In theory, quantum repeaters \cite{Briegel1998Quantum,Duan2001Long} have the potential to facilitate a global network, but the technology is still in its early stages. Recently, a fully and simultaneously connected quantum network architecture has been proposed without relying on trusted nodes \cite{Wengerowsky2018Entanglement}, which enables widespread connectivity with security guaranteed by the laws of quantum physics. Several notable entanglement-based multiuser networks were reported, showing great promise for forming scalable quantum networks \cite{Joshi2020Trusted,Liu2020Entanglement,Alshowkan2021Reconfigurable,Wen2022Realizing,Appas2021Flexible,Zheng2023Multichip,Jing2024Coexistence}.

As quantum networks approach maturity, various networked tasks have emerged, such as quantum digital signatures (QDS)\cite{Collins2014Realization,Roberts2017Experimental,Pelet2022Unconditionally,Qin2022Quantum,Yin2021Experimental}, quantum e-commerce \cite{Cao2024Experimental}, secret sharing \cite{Hillery1999Quantum,Chen2005Experimental,Williams2019Quantum}, secure anonymous protocols \cite{Hahn2020Anonymous,Huang2022Experimental}, and quantum Byzantine agreement (QBA) \cite{Fitzi2001Quantum,Iblisdir2004Byzantine,Neigovzen2008Multipartite,Gaertner2008Experimental,Rahaman2015Quantum,Smania2016Experimental,Taherkhani2018Resource,Sun2020MultiParty}. Among them, QBA, a quantum approach for handling the Byzantine generals problem, can effectively achieve consensus despite the presence of malicious players \cite{Ben2005Fast}. QBA is equipped to manage decentralized communication and computation tasks in upcoming quantum networks, applicable to various aspects of daily life like blockchain, distributed storage, distributed computation, and electronic voting \cite{Extance2015Future}. In general, it offers information-theoretic security and superior fault tolerance performance compared to classical Byzantine agreement (CBA). There are two fundamental differences between QBA and CBA protocols. One is the security loopholes of public-key encryption methods used in CBA \cite{Rivest1978Method,Castro1999Practical,Miller2016Honey}, which is under the threat of quantum computation \cite{Shor1994Algorithms,Fedorov2018Quantum}. The other lies in the 1/3 fault tolerance bound for CBA protocols, which requires a minimum of 3\textit{f}+1 players to tolerate \textit{f} malicious players \cite{Pease1980Reaching,Dolev1986Possibility}. Consequently, the three-party consensus problem is unsolvable for CBA even if the authentication classical channel is used \cite{Kiktenko2018Quantum}. The first quantum solution to the three-party consensus was proposed in 2001 \cite{Fitzi2001Quantum}, with an experimental demonstration using a four-photon entangled state in 2008 \cite{Gaertner2008Experimental}. Several relevant QBA protocols using some special entangled states or qudits were subsequently reported \cite{Iblisdir2004Byzantine,Neigovzen2008Multipartite,Rahaman2015Quantum,Smania2016Experimental}. Although these protocols offer levels of security unattainable through classical means, they deviate from Lamport’s two original Byzantine conditions by using additional assumptions. For example, the unavoidable occurrence of unreliable measurement results leads to a certain probability of failure and thus needs to be discarded. The main reason is they do not fully use the correlation to protect the unforgeability and nonrepudiation of information. Moreover, these protocols suffer from low efficiency as they require multipartite entanglement and entanglement swapping, which are unscalable and remain largely impractical due to the probabilistic nature of a spontaneous parametric down-conversion (SPDC) source. In addition, they are limited to reaching only a one-bit message consensus. Fortunately, QDS \cite{Roberts2017Experimental,Qin2022Quantum,Gottesman2001Quantum} is a potential tool to overcome all these obstacles by establishing a multiparty correlation among users. QDS can protect data integrity, authenticity, and non-repudiation with information-theoretic security. Nonetheless, there are still three main challenges of contemporary QDS related to the complexity and cost of their implementations. First, the well-known single-bit QDS protocols have a low signature rate, rendering them impractical for signing long messages. Second, most schemes require perfect keys with complete secrecy, which often entails substantial computational overhead and results in considerable delays in performing privacy amplification procedures. Third, the high demand for system manufacturability and accessibility for consumers in the network often leads to difficulty in their practical deployment.

Here, we counter the above challenges by implementing an efficient one-time universal hashing (OTUH) QDS protocol~\cite{Yin2021Experimental} on a cost-effective source-independent quantum network. By exploiting OTUH-QDS without perfect keys, we experimentally beat the limitation of 1/3 fault tolerance bound and security loopholes of CBA, indicating the quantum advantage in resolving consensus problems. This is possible owing to several recent advances. It was previously shown \cite{Li2023Onetime} that imperfect quantum keys with partial information leakage can be used for digital signatures and authentication without compromising security while having orders of magnitude improvement on signature rate compared with conventional single-bit schemes. Moreover, practical implementation of QBA not only requires a decentralized protocol design but also necessitates experimental equipment without additional trustworthiness assumptions. Intriguingly, the low-complexity fully connected quantum network architecture is an excellent experimental solution to decentralized protocols \cite{Wengerowsky2018Entanglement}, resulting in practical QBA within a strict information-theoretic secure framework. It can use an untrusted entanglement source to establish simultaneous communication between one node and all other nodes, which is different from proof-of-principle experimental demonstrations using pairwise BB84 QKD \cite{Weng2023Beating}. The last advance is the development of scalable architecture and integrated hardware for the creation of broadband bipartite entanglement, which is compatible with complementary metal-oxide semiconductor processes \cite{Wen2022Realizing,Appas2021Flexible,Zheng2023Multichip,Jing2024Coexistence}. AlGaAs semiconductor material is an outstanding platform due to its strong second-order nonlinearities, reconfigurability, and small birefringence enabling the generation of polarization-entangled states without requiring additional walk-off compensation or interferometric schemes. Moreover, as a direct-bandgap III-V compound semiconductor, AlGaAs is well suited for lasing, which paves the way for developing monolithic integration of quantum light source \cite{Boitier2014Electrically}. Integrating these ideas, we implement the protocols using an integrated AlGaAs Bragg reflection waveguide (BRW) quantum source managed by an untrusted service provider, and a time-multiplexed decoding module, composed of two unbalanced polarization maintaining interferometers (UPMIs) and a single-photon detector (SPD), held by each end user. Our monolithic source maintains polarization entanglement fidelity exceeding 94.6$\%$ across a 26-nm bandwidth, with a high brightness of 45.6 MHz/mW. Our complete entanglement distribution experiment proves the viability of practical and affordable quantum networks and identifies pathways for solving practical quantum cryptography tasks without trusted nodes. It will be of interest to a more accessible quantum internet.

\section*{RESULTS}

\begin{figure}[!htbp]
	\centering
	\includegraphics[width=115mm]{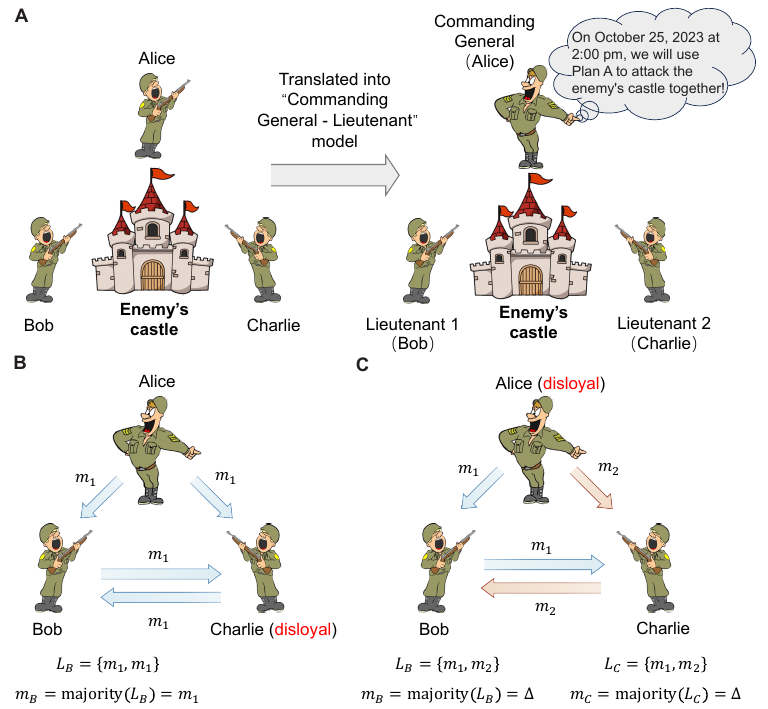}\\
	\caption{\textbf{Implementation of three-party QBA. (A)} The Byzantine generals must coordinate together to launch an attack that can overcome enemy defenses. It can be turned into a \texttt{"}commanding general-lieutenants\texttt{"} model, where the commanding general is randomly selected from among all the Byzantine generals and the others become lieutenants to reach a consensus on the commanding general’s order. \textcolor{black}{\textbf{(B)} Alice is loyal. The loyal lieutenant, Bob, can deduce the correct message $m_1$ from Alice, which satisfies IC$_2$. }\textbf{(C)} Alice is disloyal. The loyal lieutenants, Bob and Charlie, can independently deduce the consistent output message $\Delta$, which satisfies IC$_1$.
    }\label{f1}
\end{figure}

Before demonstrating the three-user QBA, we briefly review the protocol developed in \cite{Weng2023Beating}. For a strict Byzantine agreement, there are two necessary interactive consistency (IC) Byzantine conditions \cite{Lamport1982Byzantine}. The first is that all loyal lieutenants obey the same order (IC$_{1}$), and the second is that every loyal lieutenant obeys the order of the commanding general if the commanding general is loyal (IC$_{2}$). Only when both conditions are satisfied can the system reach a consensus. As shown in Fig.~\ref{f1}A, we define Alice as Commanding general, Bob as lieutenant 1, and Charlie as lieutenant 2. The main process includes three steps.

1) Alice as the signer, Bob as the forwarder, and Charlie as the verifier, they perform OTUH-QDS (see the Supplementary Materials) on the message Alice wants to send. If the signing is successful, Bob and Charlie add this valid message $m_{1}$ to their own lists $L_{B}$ and $L_{C}$, respectively.

2) Alice as the signer, Charlie as the forwarder, and Bob as the verifier (Bob and Charlie exchange the role in QDS), they perform OTUH-QDS on Alice's message again. If the signing is successful, Bob and Charlie add this valid message $m_{2}$ to their own lists $L_{B}$ and $L_{C}$, respectively.

3) Bob and Charlie will output $m_{B(C)} = {\rm majority}(m_{1}, m_{2})$ (see the Supplementary Materials) as their own final decisions.

\begin{figure}[!htbp]
	\centering
	\includegraphics[width=160mm]{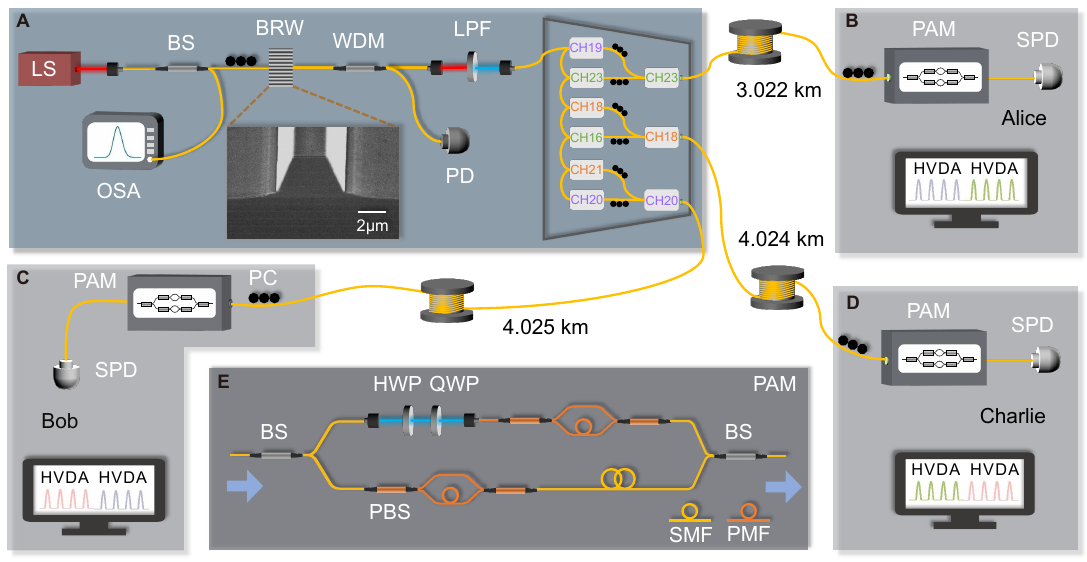}\\
	\caption{\textbf{Experimental setup, which is mainly composed of four parts. (A)} Quantum server. A 780.9-nm continuous laser is coupled into fiber and split by a 1:99 BS, where 1$\%$ of the light is injected into the optical spectrum analyzer (OSA) to monitor the laser wavelength. The remaining light is polarized and coupled into the BRW source to directly generate broadband polarization-entangled photon pairs. The inset shows the scanning electron microscopy image of the fabricated BRW sample. The source is mounted on a copper plate whose temperature is stabilized by a temperature controller. After passing through the long-pass filters, three bipartite states are selected and distributed to form a three-user fully connected network. \textcolor{black}{\textbf{(B to D)} Decoding module for three users, including PAM and one detector. }\textbf{(E)} In the PAM, each user implements basis choice by nested unbalanced interferometers and transfers polarization to photon arrival time with a single detector. Abbreviations of components: LS, laser source; WDM, wavelength division multiplexer; PC, polarization controller; PD, power detector; LPF, long-pass filter; PBS, polarization BS; HWP, half-wave plate; QWP, quarter-wave plate; PMF, polarization-maintaining fiber; SMF, single-mode fiber; SPD, single-photon detector.
    }\label{f2}
\end{figure}

To demonstrate OTUH-QDS and three-user QBA, we need to generate correlated keys between each user with the source-independent security, where entangled photon pairs are generated by an untrusted provider~\cite{Bennett1992Quantum}. We refrain from presumptions regarding the light source within the protocol, allowing the adversary to produce any state desired. However, it is essential to presume perfect Z and X bases measurements for all participants. As participants randomly select one basis for measurement, any flaws of light source inevitably result in heightened error rates, which are inherently detectable via error rate estimation. Consequently, the protocol is deemed source-independent \cite{Koashi2003Secure,Yin2020Entanglementbased}, thereby circumventing the vulnerabilities inherent in traditional prepare-and-measure protocols. The scheme of our experimental setup is shown in Fig.~\ref{f2} and includes four parts: quantum server (A) and polarization analysis module (PAM) with a detector for each user (B to D). The server is composed of an AlGaAs BRW source to prepare bipartite polarization-entangled states and a wavelength allocation unit. Owing to the dispersion and nonlinear properties of the source, the temporal walk-off between orthogonally polarized photons is so small that no compensation is required to obtain the polarization entangled state ${\rm\ket{\Psi^{+}} = \frac{1}{\sqrt{2}}(\ket{H}\ket{V}+\ket{V}\ket{H})}$, positioning it as a promising candidate for the implementation of quantum networks. The generated photon pairs can be separated into different channels using standard telecom dense wavelength division multiplexing (DWDM) filters and shared between users receiving wavelength-correlated channels due to energy conservation during the SPDC process \cite{Wengerowsky2018Entanglement}. Ultimately, we selected three pairs of channels (\{CH19, CH20\}, \{CH18, CH21\}, and \{CH16, CH23\}) to incorporate into our network architecture, with each user (that is Alice, Bob, and Charlie) receiving two channels.

\begin{figure}[!htbp]
	\centering
	\includegraphics[width=160mm]{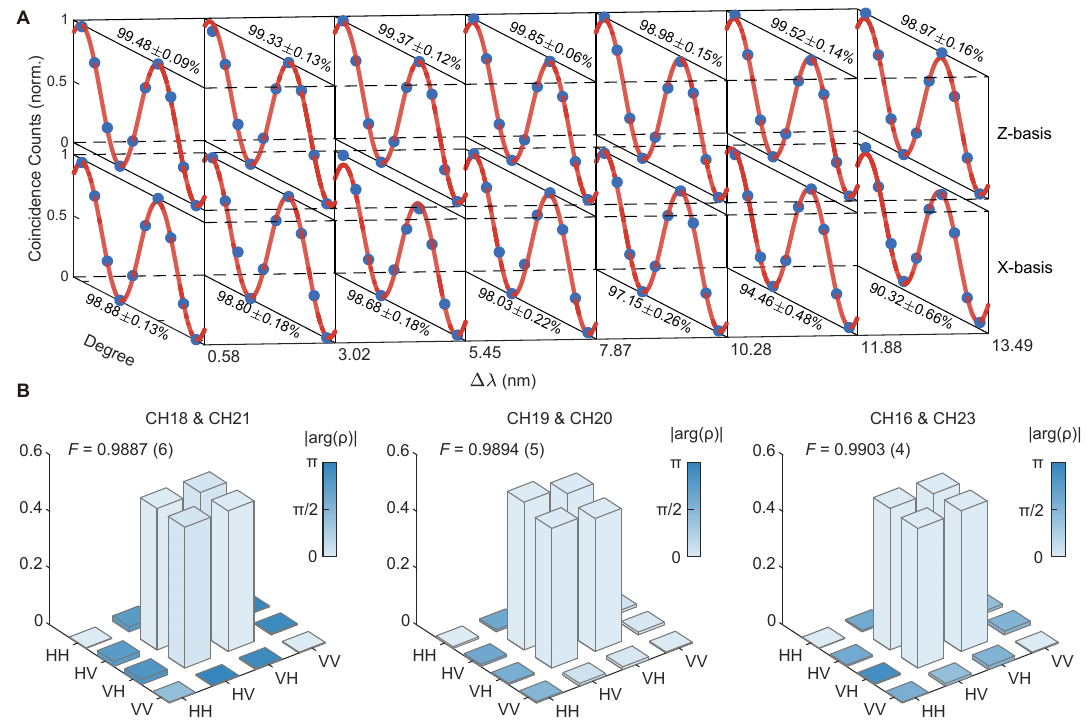}\\
	\caption{\textbf{Characterization of the BRW source. (A)} Interference curves in Z and X bases for seven different channels. The points are experimental data, and the curves are fits. The reconstructed density matrix $\rho$ of the polarization entanglement states used in the network are shown in \textcolor{black}{\textbf{(B)}. Column heights and colors represent the absolute values $|\rho|$ and phases $|{\rm arg}(\rho)|$, respectively. The uncertainties in the fidelities extracted from these density matrices are calculated using a Monte Carlo routine, assuming Poissonian errors.}
    }\label{f3}
\end{figure}

To intuitively illustrate the performance of the entangled source, we measured the interference curves as a function of the half-wave plate (HWP) angle in Z and X bases as shown in Fig.~\ref{f3}A. This measurement was conducted directly after demultiplexing but before multiplexing correlated channels to each user. The visibility of the fringe is defined as $V = ({\rm CC}_{\rm max}-{\rm CC}_{\rm min})/({\rm CC}_{\rm max}+{\rm CC}_{\rm min})$, where CC$_{\rm max}$ and CC$_{\rm min}$ are the maximum and minimum of the coincidence counts within a coincidence window of 300 ps. The fitting results show that all the raw visibilities are above 0.9, which is greater than the classical bound ($\sqrt{2}/2$), required for the violation of the Clauser-Horne-Shimony-Holt form of the Bell’s inequality. The lower bound on entanglement fidelity can be calculated by the averaged visibilities in Z (HV) and X (AD) bases \cite{Chang2016Experimental}, where H represents the horizontal polarization and V the vertical, and ${\rm\ket{D} = \frac{1}{\sqrt{2}}(\ket{H}+\ket{V})}$ and ${\rm\ket{A} = \frac{1}{\sqrt{2}}(\ket{H}-\ket{V})}$. All the fidelities can exceed 0.945 over the range. The visibility drop for large detuning with degeneracy wavelength is because of the effect of the occurrence of birefringence in the modes. To further confirm the quality of entanglement, we then perform quantum state tomography of the polarization entanglement for the frequency-correlated pairs used in the network. The photons are projected into the bases ${\rm\ket{H}/\ket{V}}$, ${\rm\ket{D}/\ket{A}}$, and ${\rm\ket{R}/\ket{L}}$, where ${\rm\ket{R} = \frac{1}{\sqrt{2}}(\ket{H}+}i{\rm\ket{V})}$ and ${\rm\ket{L} = \frac{1}{\sqrt{2}}(\ket{H}-}i{\rm\ket{V})}$. After coincidence measurements in nine different settings, the fidelity to the Bell state ${\rm\ket{\Psi^{+}} = \frac{1}{\sqrt{2}}(\ket{H}\ket{V}+\ket{V}\ket{H})}$ is estimated. Figure~\ref{f3}B displays the reconstructed density matrix of the states, showing good agreement between the maximally entangled and measured quantum states with fidelities of 98.87$\pm$0.06$\%$, 98.94$\pm$0.05$\%$, and 99.03$\pm$0.04$\%$, respectively. The maximum matrix elements of the imaginary part are smaller than 0.077. After verifying that the source can provide high-quality entanglement, we multiplexed and sent two channels to each of the three users.

\begin{figure}[!htbp]
	\centering
	\includegraphics[width=160mm]{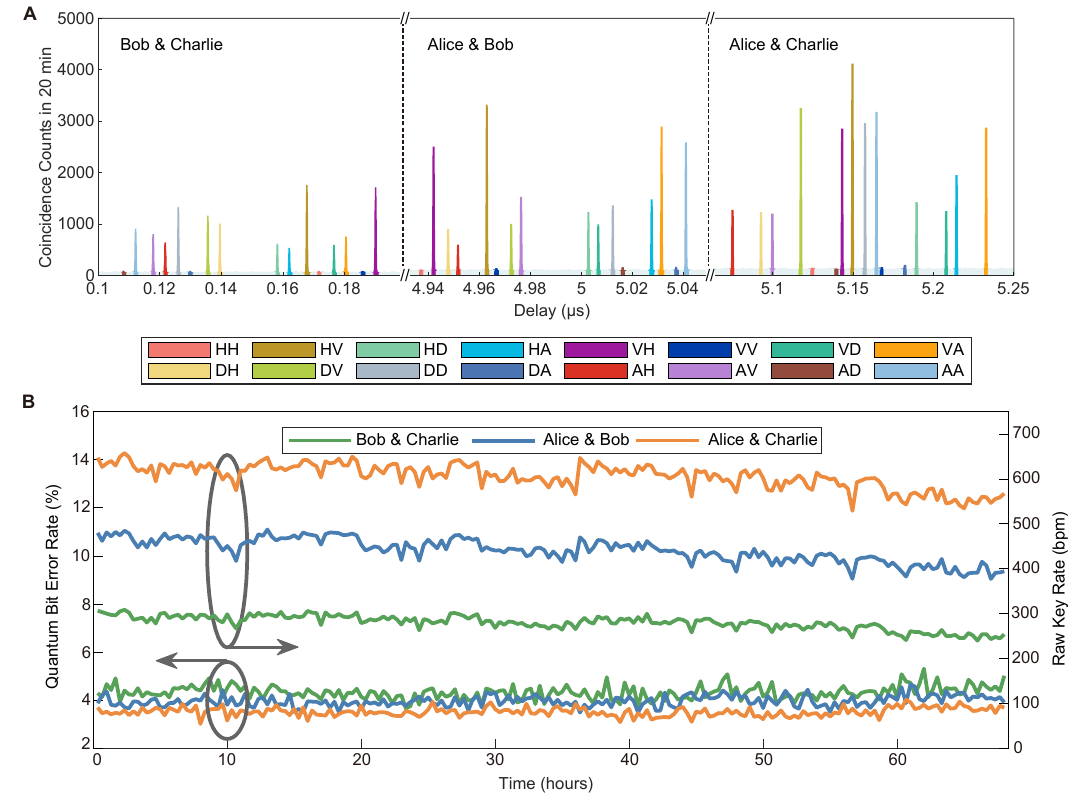}\\
	\caption{\textbf{Experimental results. (A)} Temporal cross-correlation histograms among three two-user links with 20 min of real-time data. The lower coincidence counts in the X basis are mainly due to the additional coupling losses in the analysis module. \textcolor{black}{\textbf{(B)}. Long-term performance of the network. The average $\rm{QBER = (QBER_x + QBER_z)/2}$ and sifted key rate in bits per minute (bpm) are measured for more than 68 hours.}
    }\label{f4}
\end{figure}

    To demonstrate that the bipartite states were created simultaneously in paired channels, we independently compensated all channels in two mutually unbiased bases from the source to the measurement module. Meanwhile, the multiplexing was implemented to direct two channels to each user. We connect Alice via a 3.022-km fiber spool and Bob and Charlie with 4.025- and 4.024-km spools, respectively. As shown in Fig.~\ref{f2}E, the PAM is composed of nested UPMIs (see the Supplementary Materials), in which we randomly chose the measurement basis by a 50:50 beam splitter (BS) and projected the photons to X (up) or Z (down) bases with different delays. Furthermore, entangled photon pairs were identified by their arrival time. We can consequently obtain the number of coincidence counts in 16 configurations in one data accumulation for each channel pair. Figure~\ref{f4}A shows the time-correlation histograms for the three links with a data accumulation time of 20 min. The simultaneous projection measurements on the Z and X bases are indicated by the marked 16 coincidence combinations between each pair. The four error terms ${\rm\ket{HH}}$, ${\rm\ket{VV}}$, ${\rm\ket{DA}}$ and ${\rm\ket{AD}}$ are almost submerged in accidental counts owing to the high degree of entanglement and manual optimization of the polarization. With the non-classical correlation counts measured in HV and DA bases, we calculated the lower bound on the Bell-state fidelities as 91.92$\pm$0.57$\%$ (Alice and Bob), 92.54$\pm$0.46$\%$ (Alice and Charlie) and 91.17$\pm$0.74$\%$(Bob and Charlie), respectively. These results show that we have successfully shared entanglement in all channels.

Having established high-quality bipartite states, we proceed to perform the BBM 92 protocol \cite{Bennett1992Quantum} to establish keys between each user in the network. Figure~\ref{f4}B displays the evolution of the averaged quantum bit error rate (QBER) and sifted key rates per minute in more than 68 hours of testing. Steep spikes in the QBER and key rates are mainly caused by room temperature variations, which result in disturbances to the fiber coupling efficiency. The QBERs in Z and X bases are 3.42$\%$ and 4.76$\%$ (Alice and Bob), 3.40$\%$ and 3.67$\%$ (Alice and Charlie), 3.60$\%$ and 5.43$\%$ (Bob and Charlie), respectively (see the Supplementary Materials). The corresponding averaged sifted key rates are 459.54, 628.35, and 293.07 bits per minute. After key sifting, three users reconcile their keys and adopt low-density parity-check codes for error correction. The scheme was optimized for graphics processing unit (GPU) utilization and implemented on a workstation equipped with an Intel Xeon Gold 6226R central processing unit @2.9GHz, NVIDIA GeForce RTX 3090 GPU, and 64-gigabyte RAM (random-accsee memory). By leveraging the high parallelism advantage of the GPU, the deployed reconciliation scheme achieved an average throughput exceeding 80Mbps. The average correction efficiencies are 1.1648, 1.1627, and 1.1468, respectively (see the Supplementary Materials).

\begin{figure}[!htbp]
	\centering
	\includegraphics[width=160mm]{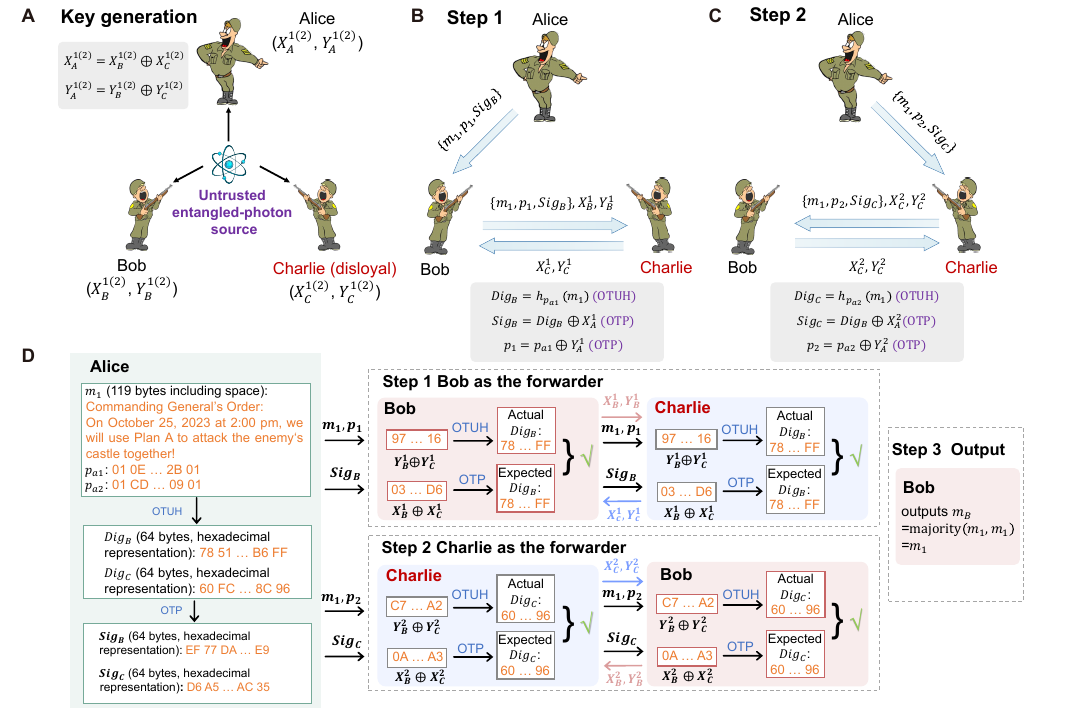}\\
	\caption{\textbf{The process of our QBA when Alice is loyal. (A)} Key generation. Through our source-independent entangled network, Alice shares four key strings $X_B^1$, $X_B^2$, $Y_B^1$, and $Y_B^2$ with Bob and four key strings $X_C^1$, $X_C^2$, $Y_C^1$, and $Y_C^2$ with Charlie, respectively. Alice obtains her key strings $X_A^1$, $X_A^2$, $Y_A^1$, $Y_A^2$ by XORing operation. \textcolor{black}{\textbf{(B)} Schematic of step 1. Alice generates $Dig_B$ through generalized division hash function decided by an irreciduble polynomial and encrypt the digest and polynomial by $X_A^1$ and $Y_A^1$ through one-time pad (OTP). Alice sends $m_1$ as well as $Sig_B$ and $p_1$ to Bob. Bob then sends $m_1$, $Sig_B$, $p_1$, and his keys $X_B^1$ and $Y_B^1$ to Charlie. Thereafter, Charlie sends his keys $X_C^1$ and $Y_C^1$ to Bob. Bob and Charlie independently recover Alice’s keys to verify the signature. \textbf{(C)} Schematic of step 2. Similar to step 1, Alice perform OTUH-QDS to sign $m_1$ by $X_A^2$  and $Y_A^2$ to Charlie. \textbf{(D)} Detailed schematic of QBA when Alice is loyal. The hash function is $h(M)=M(x)x^{64}{\rm mod}~p_{a_i}(x)$ \cite{Gottesman2001Quantum}, where $i\in \{1,2\}$ represents different steps, $p_{a_i}(x)$ is a randomly selected irreducible polynomial of order 64 in GF(256), and the coefficients of $M(x)$ correspond to every char (ASCII) of the message to be signed. The digest is $Dig=h(M)$, and the signature is $Sig_{B(C)}=Dig_{B(C)}\oplus X_A^{1(2)}$ and $p_{1(2)}=p_{a_{1(2)}}\oplus Y_A^{1(2)}$.}
    }\label{f5}
\end{figure}

Last, let us show the practical implementation of our three-user QBA. The requirement of the theory is to construct multiparty correlation of these three players to generate correlated quantum keys for three-party QDS in step 1 and 2 of QBA. As shown in Fig.~\ref{f5}A, the secret keys of Alice, Bob, and Charlie in the two steps are generated by our three-user  source-independent quantum network and satisfy the relationship $X_{A}^{1(2)} = X_{B}^{1(2)} \oplus X_{C~}^{1(2)}$ and $Y_{A}^{1(2)} = Y_{B}^{1(2)} \oplus Y_{C~}^{1(2)}$. $X_{A}^{1(2)}$ and $Y_{A}^{1(2)}$ are used for one-time pad to encrypt the hash function and digest by Alice. $X_{B}^{1(2)}$ and $Y_{B}^{1(2)}$ are owned by Bob, and  $X_{C}^{1(2)}$ and $Y_{C}^{1(2)}$ are owned by Charlie. We consider the
two situations in detail: (a) Alice is loyal, and (b) Alice is disloyal, as shown in Fig.~\ref{f1}B and ~\ref{f1}C.

(a) The commanding general Alice is loyal. In Step 1, as shown in Fig.~\ref{f5}B, Alice signs his order, $m_1$, using his keys $X_{A}^{1}$ and $Y_{A}^{1}$, and then sends the message and signature $Sig_B$ to Bob. Bob forwards them and his own secret keys $X_{B}^{1}$ and $Y_{B}^{1}$ to Charlie. After Charlie receives them, Charlie sends his keys $X_{C}^{1}$ and $Y_{C}^{1}$ to Bob. When and only when both Bob and Charlie verify the signature successfully, the signing is valid, and then Bob and Charlie add this valid message to their lists, $L_B$ and $L_C$, respectively. In Step 2, as shown in Fig.~\ref{f5}C, Charlie and Bob change their roles in QDS and follow the same process as step 1. Note that if malicious Charlie wishes that Bob would not distinguish the two messages, Charlie must forward the incorrect message $m_2$ instead of $m_1$ in step 2. However, because of the unforgeability of QDS, Charlie can only forge the message with a negligible probability $\varepsilon_{\rm{for}}$. In step 3, loyal Bob outputs his final decision, $m_B = {\rm majority}(m_1,~m_1) = m_1$, which is consistent with loyal Alice’s order and satisfies IC$_2$.

Here, we describe the OTUH-QDS without perfect keys in step 1 in detail, which is also used in step 2. Specifically, we use generalized division hash functions to generate the signature, which involve an irreducible polynomial in a Galois Field. The security parameter, i.e., the maximum probability that an attack is successfully performed in OTUH-QDS, is determined by the signature length that is usually chosen as the power of two for simplicity of encoding. The polynomial is of order $n$ in GF($2^l$), and the signature length is $nl$. We choose $l=8$ and $n=64$, so that the length of the hash value, i.e., the signature length is $512$ $(64\times 8)$, and thus the probability of forgery in our experiment, is $\varepsilon_{\rm{for}}=2.93\times10^{-16}$ which is negligible (see the Supplementary Materials).

Alice first generates a random irreducible polynomial in GF(256) of order 64 with $p_{a_1}(x)=x^{64}+x^{63}+0Ex^{62}+\ldots +2Bx+01$, where the coefficient, such as 0E and 2B, are elements in GF(256). The digest of $m_1$, i. e., $Dig_{B}$, is generated by performing a generalized division hash function decided by $p_{a_1}(x)$ on $m_1$. Before being input to the hash function, the command $m_1$ is encoded into a polynomial $m_1(x)$ by transforming every char into an element in GF(256) according to the American Standard Code for Information Interchange (ASCII) code and mapping every element into the coefficients of the polynomial in turn. The process of hashing is $h(m_1)$$=$$m_1(x)x^{64}~{\rm mod}~p_{a_1}(x)$. The output is also a polynomial and is then transformed into a string $Dig_B$ consisting of 64 elements in GF(256). Meanwhile, $p_{a_1}(x)$ is also transformed into a string $p_{a_1}$ consisting of 64 elements in GF(256). $Dig_B$ and $p_{a_1}$ are encrypted by the imperfect quantum keys through one-time pad to generate $Sig_B=Dig_B\oplus X_A^1$ and $p_1=p_{a_1}\oplus Y_A^1$.

Alice sends the commanding general’s order $m_1$ as well as $Sig_B$ and $p_1$ to Bob. Bob then sends $m_1$, $Sig_B$, $p_1$, and his keys $X_B^1$ and $Y_B^1$ to Charlie. Thereafter, Charlie sends his keys $X_C^1$ and $Y_C^1$ to Bob. After the process above, Bob and Charlie successfully share their keys. They can recover Alice’s keys $X_A^1$ and $Y_A^1$ by exclusive ORing (XORing) $X_B^1$ and $Y_B^1$ and $X_C^1$ and $Y_C^1$. Bob and Charlie independently check the signature. They will obtain the actual digest by performing the hash function derived by received $p_1$ and recovered $Y_A^1$ on the received $m_1$ and obtain the expected digest through recovered $X_A^1$ and received $Sig_B$. If the actual digest is identical to the expected, the command is valid. The details can be found in Fig.~\ref{f5}D.

\begin{figure}[!htbp]
	\centering
	\includegraphics[width=160mm]{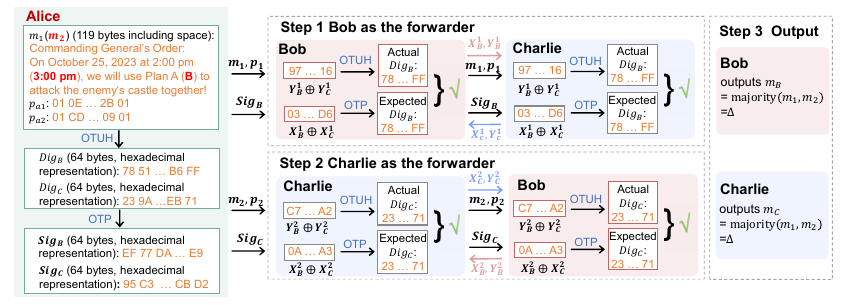}\\
	\caption{\textbf{Detailed schematic of QBA when Alice is disloyal.} The process is similar to the situation when Alice is loyal, and the difference is that disloyal Alice signs two different messages to disturb the decisions of Bob and Charlie. Here, the time of attack and the plan in $m_1$ (2:00 p.m. and Plan A) and $m_2$ (3:00 p.m. and Plan B) are different.
    }\label{f6}
\end{figure}

(b) Alice is disloyal. This situation is similar to situation (a) above. As shown in Fig.~\ref{f6}, the difference is that Alice signs different messages, $m_1$ and $m_2$, in step 1 and 2, respectively, because he wants to confuse loyal Bob and Charlie to make them output different decisions. However, because of the natural nonrepudiation of OTUH-QDS, loyal Bob and Charlie can confirm that $m_1$ and $m_2$ are both signed by Alice, and Alice is disloyal. Thus, they will output the same decision from the same information lists, i.e., $\Delta={\rm majority}(m_1,~m_2)$, where $\Delta$ is a predetermined value (see the Supplementary Materials). This satisfies IC$_1$.

\section*{DISCUSSION}
We have demonstrated the three-party QBA with imperfect keys in a fully and simultaneously connected source-independent quantum network powered by a semiconductor chip. We accomplished this by directly producing high-quality broadband polarization-entangled states with exceptional brightness using an integrated AlGaAs BRW quantum source. We realize a three-user entanglement distribution network by multiplexing six correlated DWDM channels. To reduce the financial cost and increase the scalability of such a network, we use a time-multiplexed decoding module. This module integrates two nested UPMIs and an SPD, at the expense of a slight increase in noise originating from accidentals. We implemented complete quantum communication with one detector per user in a fully connected quantum network. The combination of a cost-effective decoding module and BRW entangled source is an important step toward an affordable and practical large-scale quantum network. With the help of QKD links in our network, we generate correlated bit strings among three parties and implement OTUH-QDS with imperfect quantum keys. The signature rate can be the order of magnitude improvement owing to directly signing the hash value of long messages and the removal of cumbersome privacy amplification operations without compromising security \cite{Li2023Onetime}.

Within the framework of OTUH-QDS, we experimentally beat the 1/3 fault tolerance bound of source-independent QBA, which is unable to be achieved with classical resources. That is because OTUH-QDS can provide decentralized multiparty correlation, where all three parties participate in QDS with equal status, to remove the independence of channels with information-theoretic security, while classical digital signature schemes require a trusted third-party for signing, which disobey the decentralization of Byzantine agreement (see the Supplementary Materials). The above protocols can be used as primitives for practical multiparty quantum cryptography tasks without trusted nodes such as quantum blockchain and quantum consensus problems \cite{Fedorov2018Quantum}, paving the way for a more accessible quantum internet.

\section*{MATERIALS AND METHODS}
\noindent\textbf{BRW sample and experimental setup}

\noindent{Here we consider the properties of biphotons generated from a 7.3-mm nonideal quarter-wavelength BRW in the process of degenerate type II SPDC, i.e., one transverse electric (TE)-polarized pump photon at frequency $\omega_p$ is converted into a pair of cross-polarized signal and idler photons at frequency $\omega_s$ and $\omega_i$, respectively, with the conservation of energy $\omega_p=\omega_s+\omega_i$. The structure with a high figure of merit total nonlinearity is designed (see the Supplementary Materials), which contains a core Al$_{x_c}$Ga$_{1-x_c}$As layer with $x_c=0.17$ and a thickness of 230 nm, sandwiched between a six-period Bragg stack composed of alternative 127-nm high (Al$_{0.28}$Ga$_{0.72}$As) and 622-nm low (Al$_{0.72}$Ga$_{0.28}$As) index layers. The sample is grown along the [001] crystal axis and etched with a width of 4.8 \textmu m and a depth of 4.15 \textmu m. The waveguide achieves phase matching by using bounded total internal reflection (TIR) modes and quasi-bounded BRW modes. TIR modes are formed between high- and low-index claddings, while BRW modes are guided through transverse Bragg reflections at the interface between core and periodic claddings.}

In experiment, the BRW source is pumped by a fiber-coupled continuous wave laser centered at 780.9 nm. Lens-tapered fibers, which are used for coupling light into and out of the chip, are mounted on high-precision servo motors. We use a 980/1550 WDM to separate the pump and parametric lights, where the residual pump laser is simultaneously detected by a photodetector and acts as a feedback signal. Thereafter, the hill-climbing algorithm is adopted for real-time optimization of the coupling. The photons are further filtered by a 1500-nm long-pass filter with a high extinction ratio, which can efficiently suppress the effects of broadband photoluminescence. The photon pairs are symmetrically distributed to the degenerated wavelength. We select three pairs of correlated channels by cascaded DWDM filters with 100-GHz spacing to form a fully connected three-user network. The decoding module held by each user consists of a PAM and an InGaAs avalanche detector, implementing a passive basis choice. In the PAM, the Z basis is realized by a UPMI, where the polarization can be distinguished by the relative arrival times of photons in detectors. The X basis works alike except for an HWP set to 22.5$^{\circ}$ before the UPMI, effectively rotating the polarization by 45$^{\circ}$. The Z and X bases with different delays are combined by a  coupler (see the Supplementary Materials), and the channel is distinguished by implementing wavelength-dependent time multiplexing. The single-photon detection events are recorded using a field program-mable gate array (FPGA)-based time tag unit. Then, the information on photons’ polarization and wavelength can be transferred into their arrival time, and each user will obtain eight peaks in their temporal histogram. We used three free-running SPDs, operated at a detection efficiency of 25$\%$ and a dark count rate of 1.7 kHz with a dead time of 10 \textmu s.

\bigskip\noindent\textbf{Scalability}

\noindent{One major goal of quantum communication is to establish a network that allows for widespread connectivity, similar to the classical internet but with unconditional security based on the laws of quantum mechanics. In terms of a fully connected network, the capacity is directly proportional to the available bandwidth resource \cite{Wengerowsky2018Entanglement}. Therefore, a broader polarization-entangled photon source designed with low group birefringence \cite{Chen2019Optimizing}, closer-spaced WDM \cite{Alshowkan2021Reconfigurable}, or a combination of WDM and passive beam splitters \cite{Joshi2020Trusted} can extend the network to a larger scale. Intriguingly, multiplexing quantum sources in different spatial modes is also a promising solution to scalability, allowing high efficiency while preserving high fidelity \cite{MeyerScott2020Singlephoton}. Integrated quantum sources provide an ideal platform for implementing this scheme, which can offer stable and alignment-free operations with mature packaging technique. In the experiment, arrays of BRW samples with similar performance can be fabricated in a photonic foundry \cite{raymond2024reconfigurable}. By pumping the BRWs simultaneously, we can create $k$ subnets between $n$ users, where each BRW forms a fully connected $\frac{n}{k}$-user subnet using $\frac{n}{k}(\frac{n}{k}-1)$ wavelength channels. Each subnet can be optimized independently by adjusting the pump power to its optimal level and treated as a single user in a $k$-user network. To construct the connections between subnets, additional $k(k-1)$ wavelength channels are required and randomly distributed to all $\frac{n}{k}$ users in a subnet by passive BS. Thereby, the $n$-user network requires a total of $\frac{n}{k}(\frac{n}{k}-1)+k(k-1)$ wavelength channels. After some trivial calculation, the optimal value of $k$ is simply equal to $\sqrt{n}$. The result is valid only when $n$, $k$, $\sqrt{n}$ are all integers. Furthermore, unlike the scenario of using a single source, the source multiplexing can compensate for the beam splitting-induced losses. The unique advantage of integrating the laser directly within the AlGaAs platform can promote the network architecture in a more efficient and scalable manner \cite{Boitier2014Electrically}. After properly designing the propagation delays for photon pairs from different BRWs, the interconnection between any user in a large network can be realized without the need for a trusted node. The net effect of source multiplexing is akin to optionally constructing a two-layer fully connected network and independently controlling each subnet. One drawback of such a quantum network is the increased contribution from accidental counts resulting from the multiplexing of channels onto a single detector. It can be mitigated by shorting the coincidence window or demultiplexing the signal to multiple detectors. A pulsed pump scheme would further reduce the impact of accidental coincidence by specifying the arrival time of each channel at the detector \cite{Wengerowsky2018Entanglement}.}

The scalable source-independent fully connected network architecture is naturally suitable for the decentralized multiparty QBA scenarios. The multiparty QBA protocol can be implemented by treating three-party QDS as a basic unit and introducing a recursive structure to distribute and gather information layer by layer \cite{Weng2023Beating}. The lieutenants will output the final decision according to the information they gather in different layers. Because the basic unit of a QBA system with more than three players is three-party QDS, the full-connected network architecture can generate multiparty correlation of every three players and thus correlated quantum keys for three-party QDS. In addition, for an $N$-party system, the communication complexity, defined as the times of performing QDS, of our QBA is $C = {\sum\limits_{m = 0}^{f - 1}A_{N - 1}^{2 + m}}$, where $A_{a}^{b} = ~\frac{a~!}{(a~ - ~b)~!}$ is $b$ permutations of $a$, $f$ is the number of dishonest players, and $N$ is the number of all players ($N \geq 2f + 1$). It is evident that, as the number of players grows, the communication complexity exponentially increases. This limitation is attributed to the blockchain trilemma, which demonstrates that a decentralized system cannot achieve a harmonious balance among its essential elements: decentralization, security (fault tolerance), and scalability \cite{Zhou2020Solutions}. It is an open challenge to explore if quantum resources can break the blockchain trilemma or facilitate the relaxation of decentralization and security for designing a QBA with polynomial communication complexity.

\section*{Acknowledgments}
We thank Dr. X. Gu for helpful discussions.\newline
\textbf{Funding:}
This work was supported by the National Natural Science Foundation of China [12274233 (L.-L. L.) and 12274223 (H.-L. Y.)], the Program of Song Shan Laboratory (Included in the management of the Major Science and Technology Program of Henan Province) [221100210800-02 (H.-L. Y.)], the Innovation Program for Quantum Science and Technology [2021ZD0300700 (L.-L. L.)], and the Postgraduate Research $\&$ Practice Innovation Program of Jiangsu Province [SJCX23$\_$0569 (C. Q.)].\newline
\textbf{Author contributions:}
Conceptualization: L.-L. L., H.-L. Y., B. N., and D. J.\newline
Investigation: X. J., C. Q., C.-X. W., B.-H. L., Z. C., C.-Q. W., J. T., and X.-W. G.\newline
Visualization: X. J., C. Q., C.-X. W., and B.-H. L.\newline
Data curation: X. J., C. Q., Z. C., and J. T.\newline
Formal analysis: X. J., C.-X. W., and B.-H. L.\newline
Supervision: L.-L. L., H.-L. Y., B. N., D. J.\newline
Project administration: L.-L. L., H.-L. Y., B. N., D. J., Y.-C. K., and T.-S. C.\newline
Writing—original draft: X. J., L.-L. L., H.-L. Y., D. J., C.-X. W., and B.-H. L. \newline
Writing—review and editing: All authors.\newline
Funding acquisition: L.-L. L., H.-L. Y., and C. Q.\newline
\textbf{Competing interests:}
The authors declare that they have no competing interests.\newline
\textbf{Data and materials availability:}
All data needed to evaluate the conclusions in the paper are present in the paper and/or the Supplementary Materials.
\clearpage

\begin{center}
\fontsize{18}{0}\selectfont Supplementary Materials for\\
\fontsize{14}{40}\selectfont \textbf{Experimental Quantum Byzantine Agreement on a Three-User Quantum Network with Integrated Photonics}
\end{center}
\begin{center}
Xu Jing \emph{et al.} \\
*Corresponding author. Email: hlyin@ruc.edu.cn; jiangd@nju.edu.cn;
\\niubin$\_$1@126.com; lianglianglu@nju.edu.cn
\end{center}

\bigskip\noindent\textbf{The PDF file includes:}\\
Sections S1 to S4\\
Figs. S1 to S11\\
Tables S1 to S3\\
Reference (\emph{76-97})

\newpage

\begin{center}
\bigskip\noindent\textcolor{black}{\textbf{S1. Device and Experimental setup\\a. Device design and fabrication}}
\end{center}
The Al$_x$Ga$_{1-x}$As/GaAs Bragg reflection waveguide (BRW) structure with perfect quarter-wavelength condition has been extensively studied due to its ability to confine guided modes in the core. Here, in order to tailor the sample towards higher conversion efficiencies, we perform numerical optimization on the thickness of each layer and its aluminium concentration in the slab structures by using the finite-element method (commercial software COMSOL simulation), independent of whether the Bragg layers are perfect quarter-wavelength (QW) thick or not \cite{Yang2020Nonideal}. Depending on the specific application requirements, we define a fitness function that incorporates knowledge about key parameters, such as the type of the phase-matching (PM) processes, the mode profiles at operating wavelengths, and the group index difference of modes. The detailed flowchart of our algorithm is shown in Fig. S1.

\begin{figure}[!htbp]
	\centering
	\includegraphics[width=160mm]{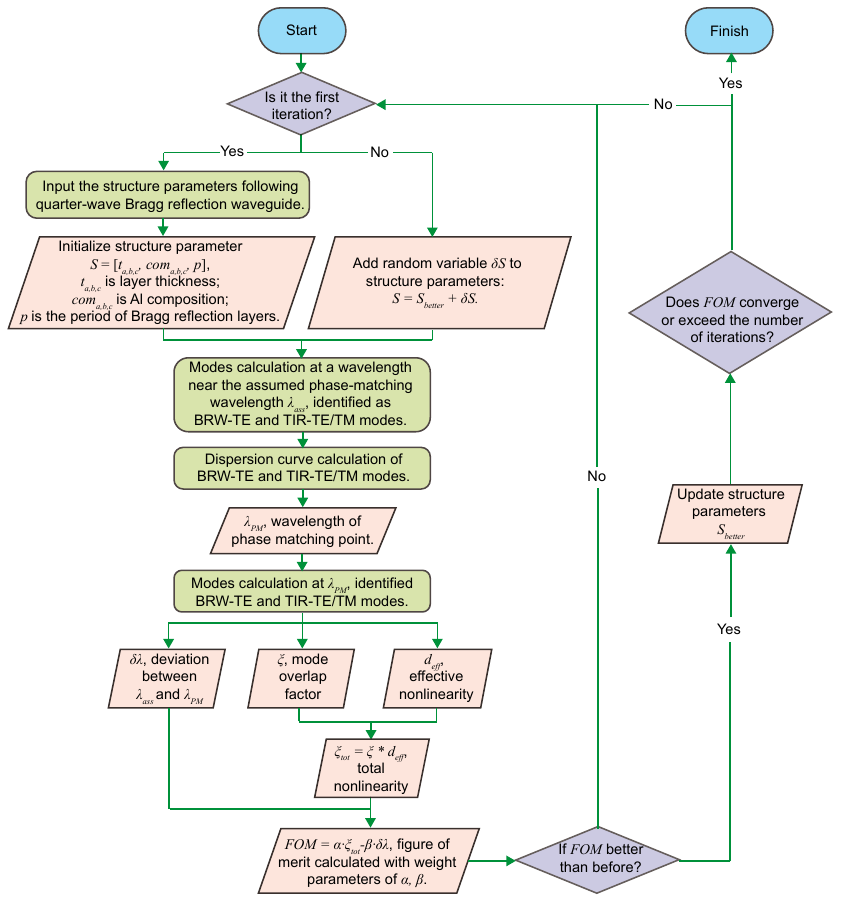}\\
	\caption*{\textbf{Fig. S1. Flowchart of the slab structural design algorithm.}
    }\label{fs1}
\end{figure}

The high modal overlap and effective nonlinearity for telecom-band degenerate type-II (${\rm TE \to TE + TM}$) PM process is set as the optimization goal. After that, the optimized structure comprises 6 cladding periods, a core Al$_x$Ga$_{1-x}$As layer with $x_c=0.13$ and thickness of 230 nm, surrounded by a Bragg stack consisting of alternating 127 nm high ($x_a=0.23$) and 622 nm low ($x_b=0.76$) index layers. However, this structure results in the BRW mode at a wavelength close to the core material room temperature bandgap, leaving little margin of error in fabrication. As there is often a discrepancy between the design and actually prepared sample. For example, due to the uncertainty of epitaxy process, there may be deviations in the thickness and composition of each layer. Impurities in the semiconductor may also be excited, leading to extra absorption loss \cite{Pressl2018Semiautomatic}. Therefore, an increase of aluminum concentration is a feasible option if samples with PM wavelengths away from the bandgap are desired \cite{Vurgaftman2001Band}. The redshift of operating wavelength can also reduce the amount of photoluminescence \cite{Auchter2021Understanding}. On the other hand, the quadratic nonlinearity of Al$_x$Ga$_{1-x}$As decreases with a reduction of the alloy composition \cite{Ohashi1993Determination}. To avoid the loss may be incurred by the uncertainty of fabrication process while maintaining a large modal overlap and effective nonlinearity, we thus respectively choose the aluminium concentration for core, high and low index layers as 0.17, 0.28 and 0.72 in fabricating our real sample. These small changes have a negligible effect on the critical performance metrics of the waveguide \cite{Schlager2021Difference}. After the slab structure was found, we carried out 2D simulations for verification, and explored the impact of the ridge widths and etching depths. The additional lateral confinement leads to a reduction in the effective mode index, which is dependent on the mode shape and wavelength \cite{Pressl2018Semiautomatic}. The simulation result shows that a 5 \textmu m wide ridge is expected to increase the PM wavelength by approximately 5.7 nm at telecom wavelengths when the waveguide is etched to the core layer. Ultimately, the structure is set with 4.8 \textmu m wide and 4.15 \textmu m deep. The normalized electric field distributions ($E_p$, $E_s$, $E_i$) along the growth direction of the proposed ridge waveguide are shown in Fig. S2. Different from the usual design, the high index in the core layer leads to a dip for BRW-TE mode at the center of the stack position. The three-mode spatial overlap can be calculated as \cite{Bijlani2012Design}
\begin{equation}\label{statflaw}\tag{S1}
	\begin{aligned}
        \xi  =\frac{\left| {{\iint{E_{p}E_{s}E_{i}dxdy}}} \right|}{\sqrt{\iint{E_{p}^2dxdy}}\sqrt{\iint{E_{s}^2dxdy}}\sqrt{\iint{E_{i}^2dxdy}}} ,
    \end{aligned}
\end{equation}
and the effective nonlinearity $d_{eff}$ is 
\begin{equation}\label{statflaw}\tag{S2}
	\begin{aligned}
        d_{eff}  = \frac{\left| {\iint{d(x,~y)E_{p}E_{s}E_{i}dxdy}} \right|}{\iint{E_{p}E_{s}E_{i}dxdy}},
    \end{aligned}
\end{equation}
where $d(x,~y)$ is the nonlinear constant for each material. The structure offers a high figure of merit total nonlinearity $\xi_{tot}$ ($d_{eff}$$\xi$), with $d_{eff}$ and $\xi$ are 72.27 pm/V and $9.86\times 10^4$ m$^{-1}$, respectively.

\begin{figure}[!htbp]
	\centering
	\includegraphics[width=120mm]{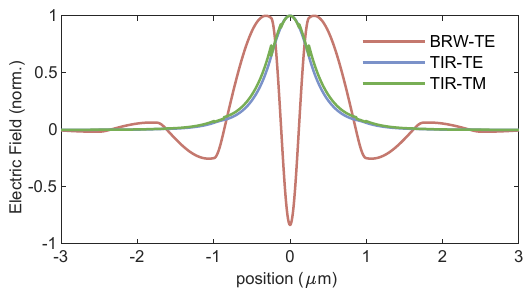}\\
	\caption*{\textbf{Fig. S2. Normalized electric field distributions along the growth direction of the proposed ridge waveguide calculated with 2D simulations.}
    }\label{fs2}
\end{figure}

The fabrication process begins with a semi-insulating GaAs wafer. The BRW and core layers are grown on the wafer by molecular beam epitaxy (MBE), in which materials are deposited one atomic layer at a time, enabling the creation of clean and sharp interfaces between different layers. During the epitaxy process, the growth rate of each layer is carefully controlled through reflection high energy electron diffraction measurement to monitor the surface and achieve high precision thickness. The ridge waveguide fabrication proceeds as follows: First, a 1 \textmu m thickness SiO$_2$ is deposited using plasma-enhanced chemical vapor deposition. Next, a layer of photoresist is applied onto the wafer, and the waveguide pattern is defined by lithography along GaAs crystal orientation [110]. Following that, the SiO$_2$ is etched away in a reactive ion etching step using fluorine gas, with the patterned photoresist serving as the etching mask. After that, the ridge waveguide is fabricated by periodic inductively coupled plasma etching using chlorine gas. To effectively reduce the solid-state reaction products, we implement in-situ nitrogen purging after each chlorine-based etching process. Finally, the sample is cut along the cleavage plane with smooth edges.

\begin{center}
\bigskip\noindent\textbf{b. Experimental setup and device characterization}\\
\end{center}
\begin{figure}[!htbp]
	\centering
	\includegraphics[width=160mm]{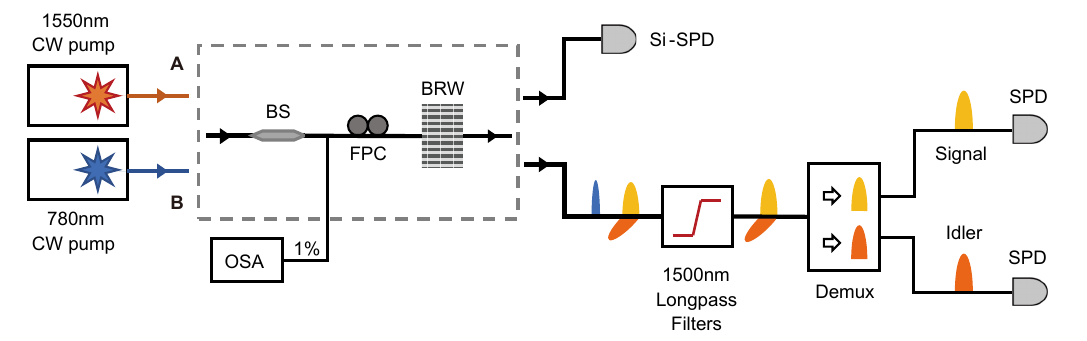}\\
	\caption*{\textbf{Fig. S3. Schematic of the experiment setup for (A) SHG and (B) spontaneous parametric down-conversion (SPDC).}
    }\label{fs3}
\end{figure}
Figure S3 depicts the experimental setups used to characterize the source. First, the PM wavelengths are determined by testing three PM processes through second harmonic generation (SHG) as shown in Fig. S3A. A tunable continuous semiconductor laser serves as the fundamental light (FL). The FL is connected to a 99:1 fiber optic beam splitter (BS), with one port (1$\%$ power) directed to the optical spectrum analyzer (OSA) for wavelength detection, and the other port (99$\%$ power) connected to the optical fiber polarization controller (FPC) before coupling into the waveguide. The SHG light is collected and directed into a Si single-photon avalanche detector. In our source, the highly versatile modal birefringence in BRW enables the PM of three modalities \cite{Abolghasem2012Monolithic,Valls2013Generation,Abolghasem2010Type} namely type-0: TM$_{\omega}$ + TM$_{\omega}$ $\to$ TM$_{2\omega}$, type-I: TE$_{\omega}$ + TE$_{\omega}$ $\to$ TM$_{2\omega}$ and type-II: TE$_{\omega}$ + TM$_{\omega}$ $\to$ TE$_{2\omega}$. By tuning the wavelength and polarization of the FL, we record the SHG as a function of FL wavelength. As shown in Fig. S4A, three peaks are observed at 1552.2 nm, 1561.8 nm and 1583.6 nm, indicating the pump wavelengths at which type-I, type-II and type-0 PM conditions are met, respectively. Additionally, we investigate the dependence of PM wavelength on temperature by varying the sample temperature from 14°C to 29°C for the type-II process. As shown in Fig. S4B, the operating wavelength is almost linear increasing from 1561.0 nm to 1563.7 nm, indicating a thermos-optic coefficient of d$\lambda$/dT = 0.18 nm/K.

\begin{figure}[!htbp]
	\centering
	\includegraphics[width=160mm]{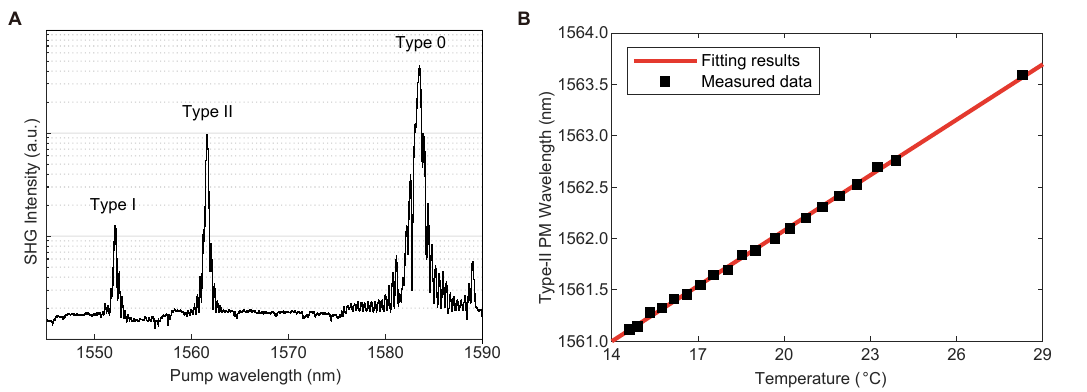}\\
	\caption*{\textbf{Fig. S4. Characterization of the SHG. (A)} The SHG intensity curves as a function of the wavelength of the fundamental light. \textbf{(B)} The type II PM wavelength as a function of sample temperature. The line is a guide to the eye.
    }\label{fs4}
\end{figure}

Next, we characterize the entangled photons by using the setup shown in Fig. S3B. A tunable laser, with a center wavelength of 780.9 nm, is polarized and coupled into the chip using lens fibers. Entangled photons emerging from the chip are filtered by 1500 nm long-pass filters and further selected by filters. The photons are then detected by two free-running single photon detectors (SPDs) with 25$\%$ detection efficiency, 1.7 kHz dark count rates, and a dead time of 10 \textmu s respectively. Then the detector signals are collected by a timetag device. To obtain the on-chip generation rate, we measure the photon counts of different polarizations, i.e., TM and TE pump, respectively. The photon counts can be expressed as
\begin{equation}\label{statflaw}\tag{S3}
	\begin{aligned}
       N_{s}^{TE} = B\eta_{s} + N_{s}^{TM},\\
       N_{i}^{TE} = B\eta_{i} + N_{i}^{TM},
    \end{aligned}
\end{equation}
where $N_{s}^{TE}$ ($N_{i}^{TE}$) and $N_{s}^{TM}$ ($N_{i}^{TM}$) represent the photon counts under different conditions of pump polarization incidence in signal (idler) channel, respectively. Here we assume the noise in the source is independent of the pump polarization. \textit{B} is the brightness of photon-pairs source in the waveguide, $\eta_{s}$ ($\eta_{i}$) represents the collection efficiency of signal (idler) photon. When the pump polarization is TM, the SPDC process does not occur. The photon counts mainly consist of the dark counts of the SPDs and the fluorescence noise generated by the materials, which can be expressed as 
\begin{equation}\label{statflaw}\tag{S4}
	\begin{aligned}
       N_{s}^{TM} = {DC}_{s} + \xi P^{\alpha}\eta_{s},\\
       N_{i}^{TM} = {DC}_{i} + \xi P^{\alpha}\eta_{i},
    \end{aligned}
\end{equation}
where ${DC}_{s}$ (${DC}_{i}$) is the dark counting rates, $\xi$ is the photoluminescence generation efficiency, \textit{P} is the pump power in front of the lens fiber coupled to the waveguide end. The exponent $\alpha$ is the coefficient related to the material. We can express the measured coincidence counts ($CC$) and accidental coincidence counts ($ACC$) as
\begin{equation}\label{statflaw}\tag{S5}
	\begin{aligned}
       CC = B\eta_{s}\eta_{i} + ACC,\\
       ACC = N_{s}N_{i}\tau_{\rm bin},
    \end{aligned}
\end{equation}
where $\tau_{\rm bin}$ = 300 ps is the temporal width of the coincidence window in our experimental parameter settings. According to the Eqs. (S3-S5), we can get the formula for calculating the source brightness as
\begin{equation}\label{statflaw}\tag{S6}
	\begin{aligned}
       B = \frac{\left( {N_{s}^{TE} - N_{s}^{TM}} \right)\left( {N_{i}^{TE} - N_{i}^{TM}} \right)}{CC - ACC}.
    \end{aligned}
\end{equation}

\begin{figure}[!htbp]
	\centering
	\includegraphics[width=80mm]{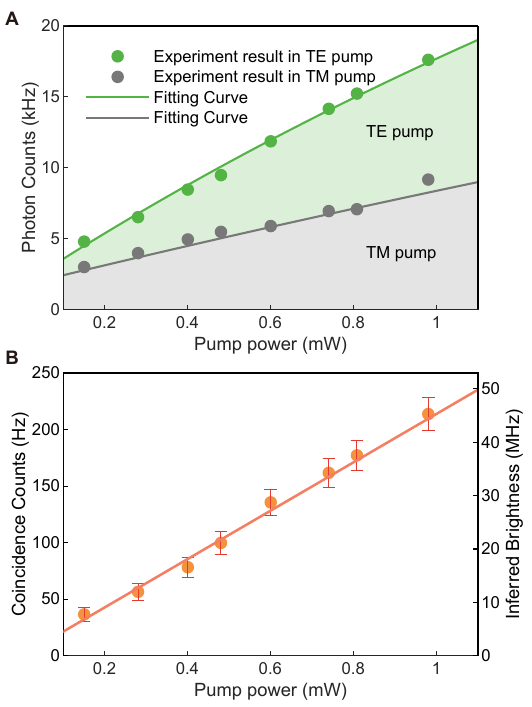}\\
	\caption*{\textbf{Fig. S5. Experimental results of generated correlated photon pairs. (A)} Photon counts in the single channel as a function of the average laser power for different pump polarization. \textbf{(B)} Measured coincidence counts and the inferred source brightness as a function of pump power.
    }\label{fs5}
\end{figure}

Figure S5A shows the photon counts for different pump polarization in the single channel as a function of pump power with 100 GHz wavelength division multiplexer. We can fit the measured data according to Eqs. (S3-S5) with the total detection efficiencies $\eta_{s}$ = $\eta_{i}$ = 0.025, which is determined by measuring the loss of light from the chip to the detector. The parameter values obtained from the fits are $\xi \approx 0.43$ MHz/mW/nm, and $\alpha \approx 1$. This indicates that the photoluminescence of the waveguide is mainly a linear process at low pump power \cite{Auchter2021Understanding}. We test the coincidence counts at different pump powers and calculate the brightness of the source by using Eq. (S6), which is shown in Fig. S5B.

\begin{figure}[!htbp]
	\centering
	\includegraphics[width=160mm]{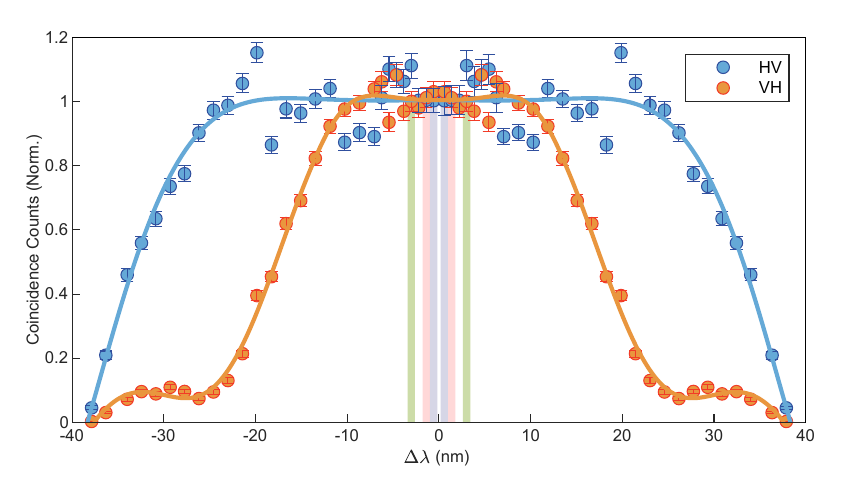}\\
	\caption*{\textbf{Fig. S6. Bandwidth characterization of the BRW source.} Experimental biphoton spectrum in H/V basis as a function of the wavelength. The channels used in the network are indicated in shades.
    }\label{fs6}
\end{figure}

To assess the effective biphoton bandwidth, we have selected symmetric channels concerning the degeneracy wavelength by using a wavelength selective switch (WSS) or coarse wavelength division multiplexer (CWDM) and a tunable filter \cite{Xu2020Secure}. For the spectral region situated in the C-band, we use a WSS, and if the region is out of the C-band we use a CWDM followed by tunable filters. In Fig. S6, we show the biphoton spectrum measured in H/V basis. The “HV” and “VH” spectrums generated exhibit notable overlap over a 13 nm range around the degeneracy point, indicating a high degree of polarization entanglement within this bandwidth, which agrees with the interference curves presented in the main text. The bandwidth can be further expanded by using shorter waveguides or designing a structure with sufficiently low group birefringence, which can be achieved through the engineering of material aluminium concentration and waveguide geometries \cite{Baboux2023Nonlinear}.

By taking into account the photon loss, bandwidth (26 nm) and 80$\%$ reflectivity of the facet for BRW mode \cite{Appas2022Nonlinear}, the generation rate can be inferred as 45.6 MHz/mW, which is an ultra-bright on-chip polarization entangled source. Table S1 compares the fully integrated polarization entangled photon-pair sources in different material platforms. It can be seen that our BRW quantum source can be considered as a leading integrated platform for generating polarization entanglement with both high brightness and broad bandwidth.

\begin{table}[ht]
	\renewcommand\arraystretch{1.2}
	\caption*{\textbf{TABLE S1. Comparison of fully polarization entanglement sources.}} 
 \resizebox{16cm}{!}{
	\begin{tabular}
		{cccccc} \hline \hline 
     Reference & Material & {\makecell[c]{Device Length\\(mm)}} & {\makecell[c]{Bandwidth\\(nm)}} & {\makecell[c]{Brightness\\(MHz/mW\\or MHz/mW$^2$)}} & {\makecell[c]{Bell-state\\Fidelity (in \%)}} \\ \hline
        \cite{Matsuda2012Monolithically} & {\makecell[c]{Silicon wire\\waveguide}} & 3 & 40 & 0.42 & 91 \\
        \cite{Chen2017Compensation} & {\makecell[c]{Periodically poled\\silica fiber}} & 185 & 90 & 1.4 & 98.7  \\
        \cite{Sun2019Compact} & Ti: LiNbO$_3$ & 11 & 2.3 & 28 & 94.5  \\
        \cite{Appas2021Flexible,Appas2022Nonlinear} & AlGaAs & 4 & 64  & 3.4  & $\textgreater$85  \\
        ~ & ~ & ~ & 26 & -- & $\textgreater$95 \\
        This work & AlGaAs & 7.3 & 26 & 45.6 & $\textgreater$94.5 \\ \hline\hline
	\end{tabular}
 }
	\label{tab1}
\end{table}

\begin{center}
\bigskip\noindent{\textbf{c. Polarization decoding module}}\\
\end{center}

\begin{figure}[!htbp]
	\centering
	\includegraphics[width=160mm]{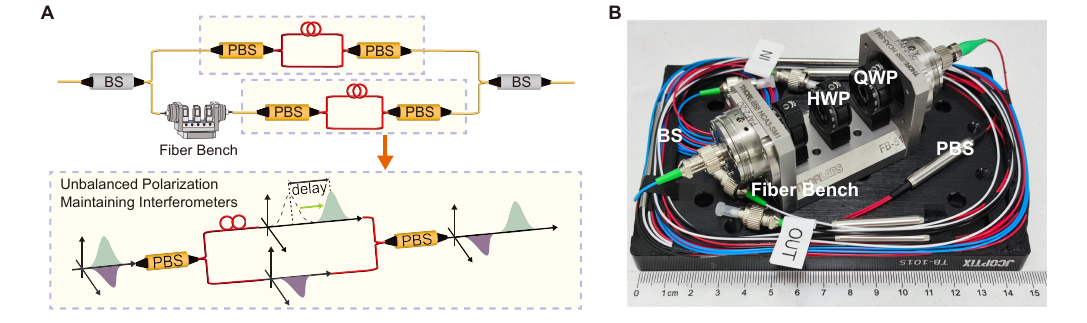}\\
	\caption*{\textbf{Fig. S7. Schematic of the compact polarization decoding module. (A)} Operating principle of the module. \textbf{(B)} Side view of the module.
    }\label{fs7}
\end{figure}

All the photons received by each user are projected onto either the horizontal/vertical (H/V) or the diagonal/antidiagonal (D/A) polarization basis. The completely passive module for implementing this projection is shown in Fig. S7A. A passive basis choice is performed using a 50:50 BS, and the basis can be distinguished by introducing a delay for detection events in the DA basis. Especially, for DA basis projection, a fiber-to-fiber U-Bench (Thorlabs PC-FFB-1550) is used for the polarization compensation and rotation, where a half-wave plate (HWP) and a quarter-wave plate (QWP) can offer full polarization control. Then the polarization-dependent shift for each basis is introduced by the fused unbalanced polarization maintaining interferometer. This enables the transfer of polarization measurement to the time when photons arrive at the detector. Figure S7B displays an image of the passive polarization decoding module developed and operational without the need for additional temperature stabilization and power consumption. The module is mounted on a 10 cm$\times$15 cm board, which is slightly larger than the packaged silicon-based decoder \cite{Wei2023Resource}. It should be noted that the second BS is used to combine the decoding information from two bases to one output, so a 3 dB loss is introduced. A fiber combiner can be used to merge two single-mode inputs into a single multi-mode output, thus avoiding this extra loss \cite{Kaltwasser2024Reducing}. Excluding the insertion losses of BS, the estimated propagation loss is less than 1.8 dB. The performances of the module are summarized in Table S2. By leveraging the off-the-shelf Fiber Bench platform and components, the module can be easily mass produced and configured for practical QKD systems. More importantly, for a large QKD network, our easy-to-handle module only requires as many detectors as connected users. This is a key step towards making QKD networks more scalable and affordable.

\begin{table}[ht]
	\renewcommand\arraystretch{1.2}
	\caption*{\textbf{TABLE S2. Propagation loss and polarization extinction ratio of the developed polarization decoding module.}} 
 \resizebox{16cm}{!}{
 \setlength{\tabcolsep}{1.4cm} 
	\begin{tabular}
		{ccc} \hline \hline 
     Channel & {\makecell[c]{Propagation loss\\(dB)}} & {\makecell[c]{Polarization extinction\\ratio (dB)}} \\ \hline
        H & 0.98(15) & $\textgreater$49 \\
        V & 0.85(22) & $\textgreater$50 \\
        D & 1.76(30) & $\textgreater$44 \\
        A & 1.68(25) & $\textgreater$47 \\ \hline\hline
	\end{tabular}
}
	\label{tab2}
\end{table}

\begin{center}
\bigskip\noindent\textbf{S2. Post-processing}
\end{center}
The quantum bit error rates (QBERs) of the sifted keys among different users are depicted as a blue line in Fig. S8, where the upper and lower parts represent the QBERs of Z and X basis, respectively. After the basis sifting process, 1758.23 kb, 2415.31 kb and 1117.12 kb sifted keys were generated between Alice and Bob, Alice and Charlie, and Bob and Charlie, respectively. To execute the error correction scheme, a set of Low-Density Parity-Check (LDPC) codes was designed. Each sifted key block, consisting of 10 kbits, underwent error correction using the appropriate LDPC code based on its estimated error rate. The error correction efficiency of sifted keys generated between different users is illustrated as a green line in Fig. S8, where the upper and lower parts represent the error correction efficiency of Z and X basis, respectively. By leveraging the high parallelism advantage of the GPU, the deployed reconciliation scheme achieved an average throughput exceeding 80 Mbps.

\begin{figure}[!htbp]
	\centering
	\includegraphics[width=160mm]{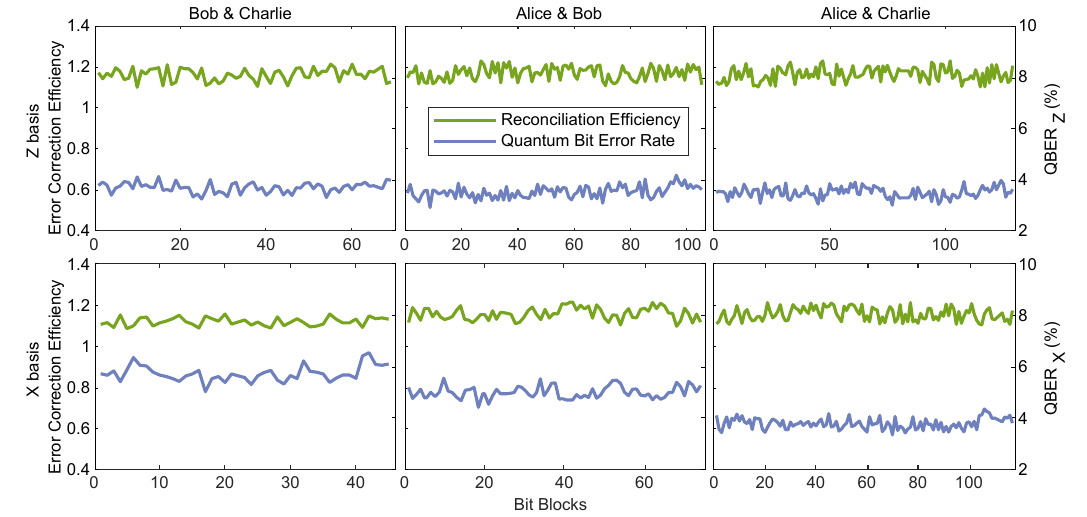}\\
	\caption*{\textbf{Fig. S8. The quantum error bit rates of the sifted keys and the error correction efficiencies for the data blocks.} Each sifted key block consists of 10 kbits.
    }\label{fs8}
\end{figure}

\begin{center}
\bigskip\noindent\textbf{S3. One-time universal hashing (OTUH) quantum digital signature (QDS) without perfect keys}
\end{center}
We use the technique of OTUH-QDS without perfect keys to perform the QDS task \cite{Li2023Onetime}. This method only requires quantum keys with full correctness and can tolerate privacy leakage. Thus, compared with QKD, the privacy amplification in the postprocessing step is removed. Here we independently give the detailed protocol of OTUH-QDS without perfect keys on the chip platform along with the security analysis as a supplementary of the main text. The protocol is consistent with that in Ref. \cite{Li2023Onetime}.

\begin{center}
\bigskip\noindent\textbf{a. Distribution stage}
\end{center}
Based on the setup in Fig. 1 and procedures illustrated above, Alice shares polarization-entangled photon pairs with Bob and Charlie, respectively. After performing measurement and encoding according to the BBM 92 protocol as depicted in the manuscript, the three parties share correlated raw keys. We choose data under X basis as the keys and those under Z basis is used for estimating privacy leakage. To perform QDS, Alice conduct error correction with Bob and Charlie on the keys under X basis, respectively. Thereafter, Alice and Bob share an identical key string denoted as $K_{B}$, and similarly Alice and Charlie share $K_{C}$. The length of $K_{B}$  ($K_{C}$) is denoted as $n_{X}^{B(C)}$, which is identical to the number of successful counts under X basis. The error correction will reveal at most $fn_{X}^{B(C)}h_2(e_{X}^{B(C)})$ bits of information, where \textit{f} is the error correction efficiency and $e_{X}^{B(C)}$  is the error rate under X basis of Alice-Bob (Alice-Charlie) in section S2. Alice obtains her string $K_{A} = K_{B} \oplus K_{C}$ by exclusive ORing (XORing) the two strings. Alice then randomly disturbs the orders of $K_{A}$, and cuts the new string into $l$-bit subgroups. Alice will publicize the new order and $l$, and Bob and Charlie will perform the same operation on $K_{B}$ and $K_{C}$ accordingly so that they also obtain several $l$-bit subgroups.

In a signature task, each of Alice, Bob, and Charlie will consume two $l$-bit strings generated above, denoted as $X_{A}$, $Y_{A}$, $X_{B}$, $Y_{B}$, $X_{C}$, $Y_{C}$, where the subscripts A, B and C represent Alice, Bob, and Charlie, respectively.

\begin{center}
\bigskip\noindent\textbf{b. Messaging stage}\\
\end{center}
\noindent 1. Alice obtains a string of random numbers through a quantum random-number generator and uses it to randomly generate a monic irreducible polynomial in GF(256) of order $k/8$, denoted as $p(x)$. $p(x)$ can be characterized through its coefficients as a bit string $p_a$.

\noindent 2. Alice transform the document to be signed $(M)$ into the coefficients of a polynomial in GF(256), denoted as $m(x)$. Alice then generates the digest through generalized division hash function $Dig = m(x)x^{64}{\rm mod}~p(x)$. She encrypts $Dig$ and $p_a$ by $X_A$ and $Y_A$, obtaining $Sig = Dig~ \oplus ~X_{A}$ and $p = p_{a} \oplus ~Y_{A}$. Alice sends $\{ M,~Sig,~p \}$ to Bob.

\noindent 3. Bob transmits $\{ M,~Sig,~p \}$ as well as $\{X_B, Y_B\}$ to Charlie, representing that he has received the signature from Bob. Thereafter, Charlie forwards his strings $\{X_C, Y_C\}$ to Bob.

\noindent 4. Bob and Charlie all recover Alice’s strings through their own and received keys. They use the recovered strings to verify the signature.

\begin{center}
\bigskip\noindent\textbf{c. Security analysis}
\end{center}
In the protocol we use imperfect keys, i.e., keys with privacy leakage, to encrypt the hash value and hash function. Thus, the security of this protocol differs from that of other QDS because of the information leakage during the distribution stage. Any possible attackers may obtain partial information on the keys. First, the attacker can obtain $fn_{X}^{B(C)}h_2(e_{X}^{B(C)})$ bits of information from the error correction process. In the following we leave out the superscript $B(C)$ for simplicity since the processes of Alice—Bob and Alice—Charlie are similar. Second, an attacker may capture the signals on the fiber channel and measure them to obtain partial keys. This part is limited by the phase error rate $e_p$ which can be estimated by the error rate under Z basis $e_Z$.
\begin{equation}\label{statflaw}\tag{S7}
	\begin{aligned}
       e_{p} = e_{Z} + \gamma( {n_{X},n_{Z},~e_{z},\varepsilon} ),
    \end{aligned}
\end{equation}
where $\gamma( {n,~k,~\lambda,\varepsilon} )$ is the statistic fluctuation function of random sampling without replacement \cite{Yin2020Tight}.
\begin{equation}\label{statflaw}\tag{S8}
	\begin{aligned}
       \gamma\left( {n,~k,~\lambda,\varepsilon} \right) = \frac{\left( {1 - 2\lambda} \right)\frac{AG}{\left( {n + k} \right)} + \sqrt{\frac{A^{2}G^{2}}{\left( {n + k} \right)^{2}} + 4\lambda\left( {1 - \lambda} \right)G}}{2 + 2\frac{A^{2}G}{\left( {n + k} \right)^{2}}}
    \end{aligned}
\end{equation}
with $A = \max\left\{ n,k \right\}$ and $G = \frac{n + k}{nk}\ln\frac{n + k}{2\pi nk\lambda\left( {1 - \lambda} \right)\varepsilon^{2}}$. For an \textit{l}-bit key string used in OTUH-QDS, i.e., a string in $\{ X_{B},Y_{B},X_{C},Y_{C}\}$, the upper bound of phase error rate is $e1_{l} = e_{p} + \gamma( {l,~n_{X}-l,~e_{p},\varepsilon} )$.

From the perspective of a possible attacker, the unknown information of an \textit{l}-bit string can be estimated by
\begin{equation}\label{statflaw}\tag{S9}
	\begin{aligned}
       {H_{l} = l\cdot\left[ {1 - h_{2}\left( {e1_{l}} \right) - fh_{2}\left( e_{X} \right)} \right]}
    \end{aligned}
\end{equation}
where $h_{2}(x) = - x\log x - \left( {1 - x} \right){\log\left( {1 - x} \right)}$. The probability that an attacker successfully guesses one of the strings $ X_{B},Y_{B},X_{C},Y_{C}$ is no more than $Pr~ = 2^{- H_{l}}$.

In a QDS scheme, Alice may be an attacker if she tries to repudiate the document she signed. Under this circumstance, Bob and Charlie are both honest. Their keys are symmetric and they will recover the same new key strings. Thus, they will make the same decision for the same document and signature. In other words, when Bob rejects (accepts) the document, Charlie also rejects (accepts) it. Therefore, our QDS protocol is immune to repudiation naturally, i.e., the repudiation bound $\varepsilon_{rep} = 0$. 

Bob can also be an attacker that he tries to forge a document that Alice did not sign and persuade Charlie to accept it. Due to the construction of hash functions, Bob’s optimal strategy is to guess the monic irreducible polynomial $p(x)$ which can be conclude if successfully guessing $X_C$. Also, since he knows that $p(x)$ is irreducible, he can guess only from all irreducible polynomials (no less than $\frac{2^{l + 2}}{l}$) rather than all polynomials ($2^l$). The probability that Bob successfully guesses $p(x)$ is $\frac{l}{4} \cdot Pr$. To tamper a $\left\{ M^{'},~Sig^{'} \right\}$ from $\left\{ M,~Sig \right\}$ that can be verified by Charlie, Bob can generate a polynomial $g(x)$ of order no more than $|M|$, and construct $M^{'}(x) = M(x) + g(x),~Sig^{'} = Sig$. As long as $p(x)$ is a factor of $g(x)$, Charlie will pass $\left\{ M^{'},~Sig^{'} \right\}$. Thus, Bob can guess $p(x)$ for no more than $\frac{|M|}{l}$ times. The final probability that Bob forge is $\varepsilon_{for} = \frac{|M|}{4}Pr = |M|2^{- 2 - H_{l}}$.

When Alice, Bob, and Charlie are all honest, the protocol will succeed unless error correction fails. Thus, the robustness bound of the protocol is $\varepsilon_{\rm rob} = 2\varepsilon_{\rm EC}$, where $\varepsilon_{\rm EC}$ is the failure probability of error correction process. The total security bound of the OTUH-QDS scheme is $\varepsilon = \max\left\{ \varepsilon_{\rm rep},\varepsilon_{\rm for},\varepsilon_{\rm rob} \right\}$.

In Fig. 5 of the main text, we show the detail of QBA and OTUH-QDS. Under this case the protocol requires Alice-Bob and Alice-Charlie channels, and the security bound is limited by Alice-Bob channel. The experimental parameters and security bound calculated by the equation above is shown in Table S3.

\begin{table}[ht]
	\renewcommand\arraystretch{1.2}
	\caption*{\textbf{Table S3. Experiment parameters of Alice-Bob channel of OUTH-QDS illustrated in Fig. 5 of the main text.}} 
 \resizebox{16cm}{!}{
 \setlength{\tabcolsep}{0.5cm} 
	\begin{tabular}
		{ccccccc} \hline \hline 
   $l$ & $|M|$ & $n_{X}$ & $n_{Z}$ & $e_{Z}$ & $e_{X}$ & $\varepsilon$ \\ \hline
       $512$ & $896$ & $1.05 \times 10^{6}$ & $7.4 \times 10^{5}$ & $0.0342$ & $0.0476$ & $2.93 \times 10^{-16}$ \\ \hline\hline
	\end{tabular}
}
	\label{tab3}
\end{table}

\begin{center}
\bigskip\noindent\textbf{d. Galois field}\\
\end{center}
Here, we add some relevant information about Galois field. Galois field, also called finite field, is a field that contains finite elements. A special Galois field is usually expressed as GF($p^n$), where $p$ is a prime number and this field contains $p^n$ elements. The elements in GF($p^n$) is usually denoted as 0, 1, 2, …, $p^n-1$. GF($p$) may be constructed as the integers modulo $p$,  i.e,
${\displaystyle \mathbb {Z} /p\mathbb {Z} }$. Calculations in GF($p$) is the same as calculations module $p$. Elements in GF($p^n$) can also be denoted as a polynomial of order $n$ that all the coefficients are elements in GF($p$). Calculations in GF($p^n$) is similar to calculations of these polynomials. The difference is that for multiplication, the outcome must be an element in GF($p^n$) which is realized by a module operation on the polynomial. The multiplicative group of GF($p^n)$ is a cyclic group of order $p^n-1$. The generalized division utilizes this character of GF($2^n$) to construct the hash function with high security.

\begin{center}
\bigskip\noindent\textbf{S4. Quantum Byzantine agreement algorithm}\\
\bigskip\noindent\textbf{a. Byzantine general problem}
\end{center}
\noindent The Byzantine General Problem (also called Byzantine fault tolerance algorithm) is a classic computer science problem that deals with the challenge of coordinating a group of distributed and autonomous entities to reach a consensus in the presence of faulty or malicious actors. In this problem, a group of Byzantine generals is camped outside a city and must coordinate their attack or retreat plans via messengers. However, some of the generals may be traitors who aim to sabotage the coordination, and messengers can be captured or corrupted during transmission, leading to false messages. 

The challenge is to design a Byzantine agreement protocol that ensures that all loyal generals agree on a common plan of action despite the existence of some disloyal generals and the risk of receiving deceitful messages. This problem has applications in distributed computing, cryptography, and especially blockchain technology. 

\begin{center}
\bigskip\noindent\textbf{b. Majority function}
\end{center}
\noindent The majority function is a logical operation that takes multiple inputs and produces an output that is the most common or majority value among its inputs \cite{Weng2023Beating}. For example, ${\rm majority}(m_{1}, \\m_{1},~m_{2}) = m_{1}$. The output of the majority function is typically biased, meaning that it will favor one of the two binary values when there is an equal number of inputs for both values. For example, ${\rm majority}(m_{1},~m_{1},~m_{2},~m_{2}) = \Delta$, where $\Delta = m_{1}$ or $m_{2}$ is a biased output which is predetermined by users. The details of quantum Byzantine agreement can be found in Ref. \cite{Weng2023Beating}.

\begin{center}
\bigskip\noindent\textbf{c. Impossibility of three-party classical Byzantine agreement}\\
\end{center}
Classical Byzantine agreements cannot break the 1/3 fault-tolerance bound, and the maximum number of malicious nodes in the system must be less than one-third of all the nodes. We take three-party scenario as an example to simply illustrate the bound. In addition, the rigorous mathematical proof for the 1/3 fault-tolerance could found in Ref. \cite{Fischer1985Easy}.

\begin{figure}[!htbp]
	\centering
	\includegraphics[width=160mm]{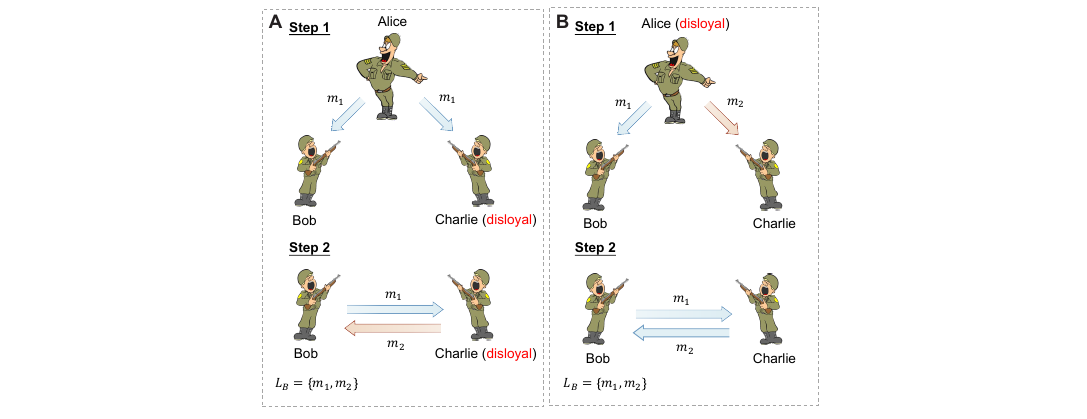}\\
	\caption*{\textbf{Fig. S9. Impossibility of three-party classical Byzantine agreement. (A)} The commanding general Alice is loyal. \textbf{(B)} The commanding general Alice is disloyal.
    }\label{fs9}
\end{figure}

In Fig. S9A, we assume that the commanding general Alice is loyal and Lieutenant Charlie is disloyal. In step 1, Alice sends his order $m_1$ to Bob and Charlie. In step 2, Bob and Charlie exchange their messages, which they received from Alice in step 1, with each other. Disloyal Charlie will replace the correct order $m_1$ with incorrect $m_2$. Thus, the information list of loyal Bob is $L_{B} = \left\{ m_{1},~m_{2} \right\}$.

In Fig. S9B, we assume that Alice is disloyal, and Lieutenants Charlie and Bob are loyal. In step 1, disloyal Alice sends different orders $m_1$ and $m_2$ to Bob and Charlie, respectively. In step 2, Bob and Charlie exchange their messages. Thus, the information list of loyal Bob is also $L_{B} = \left\{ m_{1},~m_{2} \right\}$.

We can see that because he cannot distinguish between the two above situations, loyal Bob cannot know who the traitor is, and he cannot tell what message the commander Alice actually sent to Lieutenant Charlie. Thus, he cannot deduce the output from his information list $L_B$.

\begin{center}
\bigskip\noindent\textbf{d. Two types of classical Byzantine agreements}
\end{center}
Here we introduce two types of classical Byzantine agreements, “Oral”-type and “Signature”-type, which was proposed by Lamport in 1986 \cite{Lamport1982Byzantine}. To simplify the illustration, we consider that there exists only one malicious player in both two protocols.

\noindent 1. “Oral”-type Byzantine agreement.

\begin{figure}[!htbp]
	\centering
	\includegraphics[width=160mm]{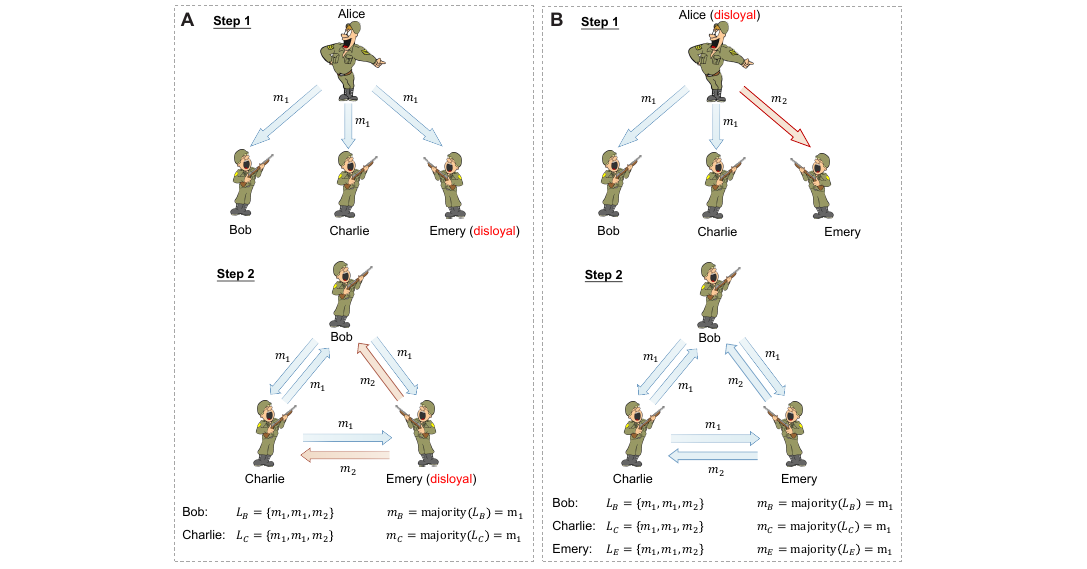}\\
	\caption*{\textbf{Fig. S10. Classical “Oral”-type Byzantine agreement. (A)} The commanding general Alice is loyal. \textbf{(B)} The commanding general Alice is disloyal.
    }\label{fs10}
\end{figure}

In Fig. S10A, we assume that the commanding general Alice is loyal, and Lieutenant Emery is disloyal. In step 1, Alice sends his order $m_1$ to Bob, Charlie and Emery. In step 2, Bob, Charlie and Emery will exchange their messages with each other, which they receive from Alice in step 1. Disloyal Emery will forge the correct order $m_1$ with incorrect $m_2$, and sends $m_2$. The information lists of loyal Bob and Charlie in step 2 are $L_{B} = \{m_{1},~m_{1},~m_{2}\}$ and $L_{C} = \{m_{1},~m_{1},~m_{2}\}$. Thus, Bob and Charlie can both deduce the correct order $m_{B} = m_{C} = {\rm majority}\{ {m_{1},~m_{1},~m_{2}} \} = m_{1}$, which satisfies IC$_2$.

In Fig. S10B, we assume that the commanding general Alice is disloyal and all the Lieutenants are loyal. In step 1, Alice sends his order $m_1$ to Bob and Charlie, $m_2$ to Emery. In step 2, Bob, Charlie and Emery will exchange their messages with each other. The information lists of loyal Bob, Charlie and Emery in step 2 are $L_{B} = \{m_{1},~m_{1},~m_{2}\}$ , $L_{C} = \{m_{1},~m_{1},~m_{2}\}$ and $L_{C} = \{m_{1},~m_{1},~m_{2}\}$. Thus, although the commanding general Alice is disloyal, Bob, Charlie and Emery can all deduce the same order $m_{B} = m_{C} = m_{E} = {\rm majority}\left\{ {m_{1},m_{1},m_{2}} \right\} = m_{1}$, which satisfies IC$_1$.

\noindent 2. “Signature”-type Byzantine agreement.

\begin{figure}[!htbp]
	\centering
	\includegraphics[width=160mm]{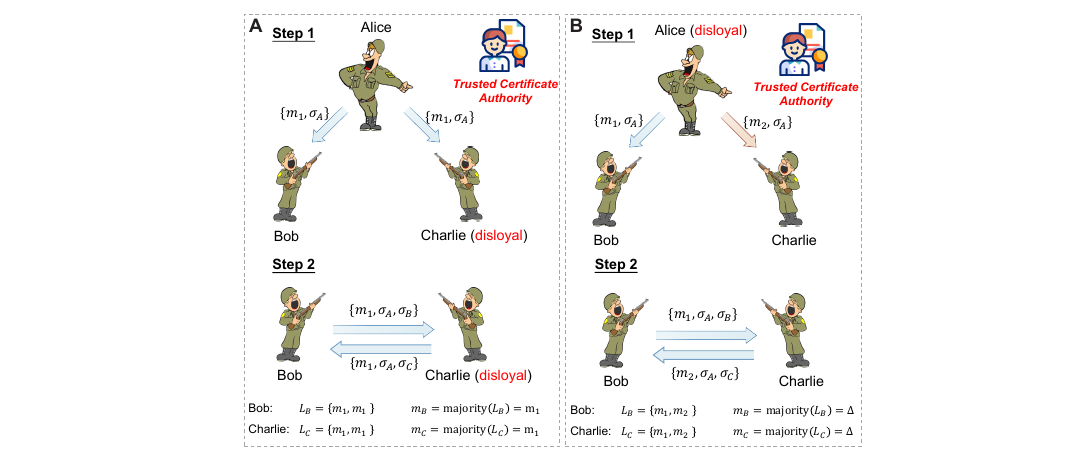}\\
	\caption*{\textbf{Fig. S11. Classical “Signature”-type Byzantine agreement. (A)} The commanding general Alice is loyal. \textbf{(B)} The commanding general Alice is disloyal.
    }\label{fs11}
\end{figure}

“Signature”-type Byzantine agreement was proposed to solve the impossibility of three-party classical Byzantine agreement by using the unforgeability and nonrepudiation of classical digital signatures. However, it does not succeed because the unavoidable existence of a trusted third party in the classical digital signatures. The trusted certificate authority sends the private key to the signer for signing, and others could use the corresponding public key to verify the signature.

In Fig. S11A, we assume that the commanding general Alice is loyal, and Lieutenant Charlie is disloyal. In step 1, Alice sends his order $m_1$ and his signature $\sigma_{A}$ to Bob and Charlie. In step 2, Bob and Charlie will exchange their messages, and attach their own signatures to their messages. Disloyal Charlie is unable to forge the correct order $m_1$ and Alice’s signature $\sigma_{A}$ due to the unforgeability of classical digital signatures. The information list of loyal Bob is $L_{B} = \{m_{1},~m_{1}\}$ and $L_{C} = \{m_{1},~m_{1}\}$. Thus, Bob can deduce the correct order $m_{B} = {\rm majority}\{{m_{1},~m_{1}}\} = m_{1}$, which satisfies IC$_2$.

In Fig. S11B, Alice is disloyal, and Charlie and Bob are loyal. In step 1, disloyal Alice sends different orders $\left\{ m_{1},~\sigma_{A} \right\}$ and $\left\{ m_{2},~\sigma_{A} \right\}$ to Bob and Charlie, respectively. In step 2, Bob and Charlie exchange their messages, $\left. \left\{ m \right._{1},~\sigma_{A},~\sigma_{B} \right\}$ and $\left\{ m_{2},~\sigma_{A},~\sigma_{C} \right\}$, with each other. Due to nonrepudiation of digital signatures, Alice cannot deny what he sends and his signature. Thus, the information list of loyal Bob and Alice are both $L_{B} = L_{C} = \left\{ m_{1},~m_{2} \right\}$. Bob and Charlie can realize that Alice is disloyal and they can negotiate a consistent output $\Delta = {\rm majority}~\left( {m_{1},~m_{2}} \right)$. 

It seems that “signature”-type Byzantine agreement achieve the three-user Byzantine agreement. However, classical digital signatures require trusted certificate authority which disobeys the decentralization of Byzantine agreement. In addition, if counting the trusted certificate authority in, there is actually four parties including one malicious player and three trusted players, which is also limited by the 1/3 fault-tolerance bound.

\begin{center}
\bigskip\noindent\textbf{e. Security analysis of three-user QBA}
\end{center}
The key to break the fault-tolerance bound is providing a decentralized multiparty correlation to remove the independence of pairwise channels \cite{Fitzi2002Detectable}. Intriguingly, quantum entanglement \cite{Fitzi2001Quantum} and asymmetric relationship of QDS \cite{Weng2023Beating} both satisfy the above requirement. QDS can provide the multiparty correlation with information-theoretical security. Up to now, we do not find classical correlations, which are greater than QDS, without any additional assumptions \cite{Wallden2015Quantum}. Almost all classical digital schemes require extra assumptions such as the existence of a trusted third party or authenticated broadcast channels \cite{Amiri2018EfficientUS}, which disobeys decentralization of Byzantine agreement.

Here we give the security analysis of our three-user QBA and show that compared with the above classical “oral” and “signature”-type Byzantine agreement, our work indeed breaks the 1/3 fault-tolerance bound due to the decentralized multiparty correlation of information-theoretically secure QDS.

According to IC$_1$ and IC$_2$, we consider whether Alice is loyal or not. \textbf{(i) Alice (commanding general) and Bob (lieutenant 1) are loyal, and Charlie (lieutenant 2) is disloyal.} Since the commanding general, Alice, is loyal, the loyal lieutenant, Bob, must obey the correct order $m_1$ from Alice. In Step 1 and Step 2, Alice sends the correct message $m_1$. Bob records $m_1$ when he acts as a forwarder in Step 1. Bob also records $m_1$ forwarded by Charlie when Bob acts as the verifier in Step 2, because malicious Charlie has to honestly forward $m_1$ due to the unforgeability of QDS. The information list of loyal Bob is $L_{B} = \{m_{1},~m_{1}\}$. The final output of loyal Bob is $m_{B} = {\rm majority}( m_{1},~m_{1}) = m_1$, which is consistent with the message sent by loyal Alice. That satisfies IC$_2$. \textbf{(ii) Alice is disloyal, and Bob and Charlie are loyal.} Since the commanding general, Alice, is disloyal, the loyal lieutenants, Bob and Charlie, are required to reach an agreement over the conflicting messages from Alice. In Step 1 and Step 2, Alice signs and sends conflicting messages $m_1$ and $m_2$ to loyal lieutenants, respectively.The information lists of Bob and Charlie are both $L_{B} = L_{C} = \{m_{1},~m_{2}\}$, and thus their outputs are both $m_{B} = m_{C} = ~{\rm majority}\left( m_{1},~m_{2} \right) = \Delta$, where $\Delta$ is a biased output of the majority function. Due to the nonrepudiation of QDS, Alice cannot deny her authority of the signatures. In addition, due to the unforgeability of QDS, no one can tamper with the messages and signatures. Based on the above two properties of QDS, Loyal Bob and Charlie can deduce that both of the two messages definitely come from Alice and Alice is disloyal. Bob and Charlie can trust each other and, for example, set the biased output as the message forwarded by Bob, i.e., $\Delta = m_1$. Although Alice is disloyal, loyal Bob and Charlie still output the same biased order $\Delta$, which satisfies IC$_1$.

\end{sloppypar}
\end{document}